\documentclass[twocolumn]{aastex63}
\usepackage{graphicx}
\usepackage[countmax]{subfloat}
\usepackage{booktabs}
\usepackage{mwe}
\usepackage{amsmath}
\newcommand{\RNum}[1]{\uppercase\expandafter{\romannumeral #1\relax}}

\begin{document}
\title{Optical Variability of the TeV Blazar 1ES 0806$+$524 on Diverse Timescales}

\correspondingauthor{Ashwani Pandey}
\email{ashwanitapan@gmail.com}

\author{Ashwani Pandey}
\affiliation{Aryabhatta Research Institute of Observational Sciences (ARIES), Manora Peak, Nainital 263002, India}
\affiliation{Department of Physics, DDU Gorakhpur University, Gorakhpur 273009, India}

\author{Alok C. Gupta} 
\affiliation{Aryabhatta Research Institute of Observational Sciences (ARIES), Manora Peak, Nainital 263002, India}

\author{Sofia O. Kurtanidze} 
\affiliation{Abastumani Observatory, Mt. Kanobili, 0301 Abastumani, Georgia}
\affiliation{Zentrum f\"ur Astronomie der Universit\"at Heidelberg, Landessternwarte, K\"onigstuhl 12, 69117 Heidelberg, Germany}
\affiliation{Samtskhe-Javakheti State University, Rustaveli Str. 113, 0080 Akhaltsikhe, Georgia}

\author{Paul J. Wiita}
\affiliation{Department of Physics, The College of New Jersey, 2000 Pennington Rd., Ewing, NJ 08628-0718, USA}

\author{G. Damljanovic}
\affiliation{Astronomical Observatory, Volgina 7, 11060 Belgrade, Serbia}

\author{R. Bachev} 
\affiliation{Institute of Astronomy and National Astronomical Observatory, Bulgarian Academy of Sciences, 72 Tsarigradsko Shosse Boulevard, 1784 Sofia, Bulgaria}

\author{Jin Zhang}
\affiliation{Key Laboratory of Space Astronomy and Technology, National Astronomical Observatories, Chinese Academy of Sciences, Beijing 100012, Peopleʼs Republic of China}

\author{O. M. Kurtanidze} 
\affiliation{Abastumani Observatory, Mt. Kanobili, 0301 Abastumani, Georgia}
\affiliation{Zentrum f\"ur Astronomie der Universit\"at Heidelberg, Landessternwarte, K\"onigstuhl 12, 69117 Heidelberg, Germany}
\affiliation{Engelhardt Astronomical Observatory, Kazan Federal University, Tatarstan, Russia}
\affiliation{Center for Astrophysics, Guangzhou University, Guangzhou 510006, People's Republic of China}

\author{A. Darriba} 
\affiliation{American Association of Variable Star Observers (AAVSO), 49 Bay State Road, Cambridge, MA 02138, USA}
\affiliation{Group M1, Centro Astron\'omico de Avila, Madrid, Spain}

\author{R. A. Chigladze}
\affiliation{Abastumani Observatory, Mt. Kanobili, 0301 Abastumani, Georgia}
\affiliation{Samtskhe-Javakheti State University, Rustaveli Str. 113, 0080 Akhaltsikhe, Georgia}

\author{G. Latev} 
\affiliation{Institute of Astronomy and National Astronomical Observatory, Bulgarian Academy of Sciences, 72 Tsarigradsko Shosse Boulevard, 1784 Sofia, Bulgaria}

\author{M. G. Nikolashvili}
\affiliation{Abastumani Observatory, Mt. Kanobili, 0301 Abastumani, Georgia}
\affiliation{Zentrum f\"ur Astronomie der Universit\"at Heidelberg, Landessternwarte, K\"onigstuhl 12, 69117 Heidelberg, Germany}

\author{S. Peneva}
\affiliation{Institute of Astronomy and National Astronomical Observatory, Bulgarian Academy of Sciences, 72 Tsarigradsko Shosse Boulevard, 1784 Sofia, Bulgaria}

\author{E. Semkov}
\affiliation{Institute of Astronomy and National Astronomical Observatory, Bulgarian Academy of Sciences, 72 Tsarigradsko Shosse Boulevard, 1784 Sofia, Bulgaria}

\author{A. Strigachev}
\affiliation{Institute of Astronomy and National Astronomical Observatory, Bulgarian Academy of Sciences, 72 Tsarigradsko Shosse Boulevard, 1784 Sofia, Bulgaria}

\author{S. N. Tiwari}
\affiliation{Department of Physics, DDU Gorakhpur University, Gorakhpur 273009, India}

\author{O. Vince}
\affiliation{Astronomical Observatory, Volgina 7, 11060 Belgrade, Serbia}


\begin{abstract}
We report the results of our optical (VRI) photometric observations of the TeV blazar 1ES 0806$+$524 on 153 nights during 2011--2019 using seven optical telescopes in Europe and Asia. We investigated the variability of the blazar on intraday as well as on long-term timescales. We examined eighteen intraday light curves for flux and color variations using the most reliable power-enhanced {\it F}-test and the nested ANOVA test. Only on one night was a small, but significant, variation found, in both $V$ band and $R$ band light curves. The $V-R$ color index was constant on every one of those nights.  Flux density changes of around 80\% were seen over the course of these eight years in multiple bands.
We found a weighted mean optical spectral index of 0.639$\pm$0.002 during our monitoring period by fitting a power law ($F_{\nu} \propto \nu^{-\alpha}$) in 23 optical ($VRI$) spectral energy distributions of 1ES 0806$+$524. We discuss different possible mechanisms responsible for blazar variability on diverse timescales.
\end{abstract}


\section{Introduction} \label{sec:intro}
Active galactic nuclei (AGNs) are universally believed to be powered by accreting supermassive black holes ($\geq 10^6$ M$_{\sun}$) lying at their centers \citep{1984ARA&A..22..471R}. About 10$-$15\% of AGNs clearly include two well-collimated jets of relativistic particles and are known as jetted-AGNs \citep{2017A&ARv..25....2P}. These relativistic jets are particularly luminous at radio and $\gamma-$ray frequencies. The different jetted-AGNs are distinguished by their viewing angles with respect to the observer's line of sight, with blazars being those in which the relativistic jets are aligned at very small ($\leq$ 15-20$^{\circ}$) viewing angles \citep{1995PASP..107..803U}. Blazars are further divided into BL Lacertae objects (BL Lacs; EW\footnote{rest frame equivalent width}$_{rest}$ $< 5$\AA) and flat-spectrum radio quasars (FSRQs; EW$_{rest}$ $> 5$\AA) on the basis of the strength of emission lines in their optical/ultraviolet (UV) spectra \citep[e.g.][]{1991ApJS...76..813S, 1996MNRAS.281..425M}. Their nonthermal broadband spectral energy distributions (SEDs) have a characteristic double bump structures \citep[e.g.][]{1998MNRAS.299..433F}. The low-frequency component, which peaks in infrared to X-ray frequencies, is well interpreted as synchrotron emission from the relativistic electrons in the jet. Based on the rest frame synchrotron peak frequency ($\nu_{peak}^S$), \cite{2010ApJ...716...30A} classified blazars as low synchrotron peaked blazars (LSPs; $\nu^S_{peak} \leq 10^{14}$ Hz), intermediate synchrotron peaked blazars (ISPs; $10^{14}$ Hz $< \nu^S_{peak} < 10^{15}$Hz), and high synchrotron peaked blazars (HSPs; $\nu^S_{peak} \geq 10^{15}$Hz). The high-frequency component peaks at GeV energies in LSPs (including FSRQs and low-frequency peaked BL Lacs (LBLs)) and at TeV energies in HSPs or high-frequency peaked BL Lacs (HBLs). The physical origin of the high-frequency component is still debated \citep[e.g.][]{2007Ap&SS.307...69B}.

Blazars are known for their highly variable nature on diverse timescales. On the basis of variability timescales ($t_{var}$), blazar variability can be broadly classified as: intraday variability (IDV) or microvariability ($t_{var} \sim$ less than a day), short-term variability (STV; $t_{var} \sim$ few days to few months) and long-term variability (LTV; $t_{var} \sim$ several months to years). Optical variability in the light curves of blazars has been studied extensively on all these three timescales by several authors \citep[e.g.,][and references therein]{1989Natur.337..627M, 1990PhDT........11C, 1995ARA&A..33..163W, 2004A&A...422..505G, 2016MNRAS.458.1127G}. Optical observations of blazars reveal that color changes are often observed together with flux variations. BL Lacs are found to follow a bluer-when-brighter (BWB) color trend, while a redder-when-brighter (RWB) color behavior is usually seen in FSRQs \citep[e.g.,][]{2012AJ....143..108W, 2015A&A...573A..69W}.

The AGN 1ES 0806$+$524 ($\alpha_{\rm 2000} = 08^h09^m49.2^s$; $\delta_{\rm 2000} = +52^{\circ}18^{\prime}58.4^{\prime\prime}$; z = 0.138 \citep{1998A&A...334..459B}) was classified as a BL Lac object on the basis of its featureless optical spectra \citep{1993ApJ...412..541S}. It was first detected at very high energy (VHE, E $>$ 300 GeV) $\gamma-$ray in 2008 by the Very Energetic Radiation Imaging Telescope Array System (VERITAS) \citep{2009ApJ...690L.126A}. 
Optical $R-$band observations of 1ES 0806$+$524 were carried out from 1997 December 27 to 2006 May 30 by \cite{2009ASPC..408..266K}. During this period it was detected in the brightest state of $R=14.81$ on 2004 September 13, while the faintest magnitude observed was $R=15.69$ on 1999 on April 7 and the maximum brightness variation was $\Delta R=0.88$.
\cite{2011MNRAS.416..101G} monitored this blazar on 4 nights but didn't find IDV on any night. Quasi-simultaneous multicolor observations of 1ES 0806$+$524 were performed from 2005 December to 2011 February by \cite{2014AJ....148..110M} and they found a strong BWB trend on long timescales, but no variation on IDV timescales.

\begin{table*}
\caption{Details of telescopes and instruments used} 
\label{tab:teles} 
\centering 
\resizebox{\textwidth} {!}{ 
\begin{tabular}{lccccccc} 
\hline\hline 
Code & A & B & C & D & E & F & G \\ 
\hline 
Telescope & 1.3 m DFOT & 1.04 m ST & 60 cm AO & 1.4 m ASV & 70 cm & 50/70 cm NAO & 35.6 cm C14 XLT\\ 
CCD Model & Andor 2K & PyLoN & FLI PL9000 & Andor iKon-L & Apogee Ap6E & FLI PL16803 & ATIK 383L+ Monochrome\\
Chip Size (pixels) & 2048 $\times$ 2048 & 1340 $\times$ 1300 & 3056 $\times$ 3056 & 2048 $\times$ 2048 & 1024 $\times$ 1024 & 4096 $\times$ 4096 & 3354 $\times$ 2529\\
Scale (arcsec/pixel) & 0.535 & 0.305 & 1.0 & 0.244 & 2.4 & 1.079 & 1.38 \\ 
Field ($arcmin^2$) & 18 $\times$ 18 & 6.8 $\times$ 6.6 & 17 $\times$ 17 & 8.3 $\times$ 8.3 & 15 $\times$ 15 & 73.66 $\times$ 73.66 & 25.46 $\times$ 19.16\\ 
Gain ($e^-$/ADU) & 2.0 & 4.0 & 1.0 & 1.0 & 8.0 & 1.0 & 1.8\\ 
Read-out Noise ($e^-$ rms) & 7.0 & 6.4 & 9.0 & 7.0 & 14.0 & 9.0 & 7.0\\ 
Typical Seeing (arcsec) & 1.2-2.0 & 1.2-2.1 & 2.0-3.0 & 1.0-1.5 & 1.0-2.0 & 2.0-4.0 & 1.3-2.0\\ 
\hline 
\end{tabular}
}
\begin{flushleft}
A: 1.3 m Devasthal Fast Optical Telescope (DFOT) at ARIES, Nainital, India. \\
B: 1.04 m Sampuranand Telescope (ST) at ARIES, Nainital, India.\\
C: 60 cm Cassegrain telescope at Astronomical Observatory (AO) Belogradchik, Bulgaria.\\
D: 1.4 m telescope at Astronomical Station Vidojevica (ASV), Serbia. \\
E: 70 cm meniscus telescope at Abastumani Observatory, Georgia.\\
F: 50/70 cm Schmidt telescope at National Astronomical Observatory (NAO), Rozhen, Bulgaria.\\
G: 35.6 cm Celestron C14 XLT telescope at Las Casqueras, Spain.
\end{flushleft}
\end{table*}

\begin{table*}
\caption{Observation log for the TeV blazar 1ES 0806$+$524.} 
\label{tab:obs_log} 
\centering
\resizebox{\textwidth} {!}{ 
\begin{tabular}{lcccccccc} 
\hline\hline 
Observation date & Telescope & Data points & Observation date & Telescope & Data points & Observation date & Telescope & Data points \\
dd$-$mm$-$yyyy & & ~V,~R,I & dd$-$mm$-$yyyy & & ~V,~R,I & dd$-$mm$-$yyyy & & ~V,~R,I \\
\hline
25$-$01$-$2011   &    C       & ~1,~1,1 &	24$-$02$-$2017   &    G       & ~1,~0,0 &	17$-$01$-$2018   &    E       & ~0,~4,0 \\ 
06$-$02$-$2011   &    C       & ~1,~1,1 &	25$-$02$-$2017   &    G       & ~1,~0,0 &	20$-$01$-$2018   &    G       & ~1,~0,0 \\
07$-$02$-$2011   &    C       & ~1,~1,1 &	28$-$02$-$2017   &    E       & ~0,~4,0 &	27$-$01$-$2018   &    G       & ~1,~0,0 \\
27$-$05$-$2011   &    C       & ~1,~1,1 &	05$-$03$-$2017   &    E       & ~0,~4,0 &	01$-$02$-$2018   &    E       & ~0,~3,0 \\
10$-$04$-$2012   &    C       & ~1,~1,1 &	10$-$03$-$2017   &    G       & ~1,~0,0 &	02$-$02$-$2018   &    G       & ~1,~0,0 \\
12$-$04$-$2012   &    C       & ~1,~1,1 &	11$-$03$-$2017   &    G       & ~1,~0,0 &	08$-$02$-$2018   &    B       & 25,25,1 \\
12$-$05$-$2012   &    C       & ~1,~1,1 &	14$-$03$-$2017   &    E       & ~0,~4,0 &	09$-$02$-$2018   &    B       & 25,25,1 \\
12$-$02$-$2015   &    C       & ~1,~1,1 &	17$-$03$-$2017   &    G       & ~1,~0,0 &	10$-$02$-$2018   &    G       & ~1,~0,0 \\
13$-$02$-$2015   &    C       & ~1,~1,1 &	18$-$03$-$2017   &    G       & ~1,~0,0 &	14$-$02$-$2018   &    E       & ~0,~4,0 \\
19$-$04$-$2015   &    C       & ~0,~1,1 &	19$-$03$-$2017   &    E       & ~0,~4,0 &	17$-$02$-$2018   &    G       & ~1,~0,0 \\
22$-$04$-$2015   &    C       & ~0,~1,1 &	19$-$03$-$2017   &    G       & ~1,~0,0 &	21$-$02$-$2018   &    A       & 30,30,1 \\
25$-$04$-$2015   &    C       & ~0,~1,1 &	28$-$03$-$2017   &    E       & ~0,~5,0 &	24$-$02$-$2018   &    G       & ~1,~0,0 \\
14$-$12$-$2015   &    C       & ~0,~1,1 &  	01$-$04$-$2017   &    E       & ~0,~4,0 &	25$-$02$-$2018   &    G       & ~1,~0,0 \\
18$-$01$-$2016   &    E       & ~0,~3,0 & 	01$-$04$-$2017   &    G       & ~1,~0,0 &	26$-$02$-$2018   &    E       & ~0,~3,0 \\
02$-$02$-$2016   &    E       & ~0,~4,0 &	07$-$04$-$2017   &    G       & ~1,~0,0 &	10$-$03$-$2018   &    G       & ~1,~0,0 \\
03$-$02$-$2016   &    E       & ~0,~4,0 &	08$-$04$-$2017   &    G       & ~1,~0,0 &	12$-$03$-$2018   &    E       & ~0,~7,0 \\
19$-$02$-$2016   &    E       & ~0,~6,0 &	10$-$04$-$2017   &    E       & ~0,~5,0 &	24$-$03$-$2018   &    G       & ~1,~0,0 \\
29$-$02$-$2016   &    E       & ~0,~4,0 &	15$-$04$-$2017   &    E       & ~0,~6,0 &	31$-$03$-$2018   &    G       & ~1,~0,0 \\
18$-$03$-$2016   &    E       & ~0,~4,0 &	21$-$04$-$2017   &    G       & ~1,~0,0 &	04$-$04$-$2018   &    E       & ~0,~4,0 \\
30$-$03$-$2016   &    E       & ~0,~6,0 &	25$-$04$-$2017   &    E       & ~0,~4,0 &	09$-$04$-$2018   &    E       & ~0,~4,0 \\
03$-$04$-$2016   &    E       & ~0,~4,0 &	28$-$04$-$2017   &    G       & ~1,~0,0 &	14$-$04$-$2018   &    G       & ~1,~0,0 \\
11$-$04$-$2016   &    E       & ~0,~4,0 &	30$-$04$-$2017   &    E       & ~0,~5,0 &	15$-$04$-$2018   &    E       & ~0,~4,0 \\
17$-$04$-$2016   &    E       & ~0,~8,0 &	12$-$05$-$2017   &    E       & ~0,~4,0 &	23$-$04$-$2018   &    E       & ~0,~4,0 \\
01$-$05$-$2016   &    E       & ~0,~4,0 &	22$-$05$-$2017   &    E       & ~0,~8,0 &	29$-$04$-$2018   &    G       & ~1,~0,0 \\
19$-$08$-$2016   &    G       & ~1,~0,0 &	26$-$05$-$2017   &    G       & ~1,~0,0 &	30$-$04$-$2018   &    G       & ~1,~0,0 \\
23$-$08$-$2016   &    E       & ~0,~2,0 &	08$-$06$-$2017   &    E       & ~0,~3,0 &	04$-$05$-$2018   &    G       & ~1,~0,0 \\
30$-$08$-$2016   &    E       & ~0,~4,0 &	09$-$06$-$2017   &    G       & ~1,~0,0 &	12$-$05$-$2018   &    G       & ~1,~0,0 \\
18$-$09$-$2016   &    E       & ~0,~2,0 &	16$-$06$-$2017   &    G       & ~1,~0,0 &	14$-$05$-$2018   &    G       & ~1,~0,0 \\
02$-$10$-$2016   &    E       & ~0,~3,0 &	23$-$06$-$2017   &    E       & ~0,~4,0 &	15$-$06$-$2018   &    G       & ~1,~0,0 \\
03$-$10$-$2016   &    E       & ~0,~5,0 &	03$-$07$-$2017   &    E       & ~0,~5,0 &	17$-$11$-$2018   &    G       & ~1,~0,0 \\
12$-$10$-$2016   &    E       & ~0,~5,0 &	29$-$08$-$2017   &    E       & ~0,~4,0 &	02$-$12$-$2018   &    D       & ~3,~3,3 \\
08$-$11$-$2016   &    E       & ~0,~5,0 &	04$-$09$-$2017   &    E       & ~0,~4,0 &	15$-$12$-$2018   &    A       & 30,30,1 \\
13$-$11$-$2016   &    E       & ~0,~6,0 &	12$-$09$-$2017   &    E       & ~0,~4,0 &	16$-$12$-$2018   &    A       & 41,41,1 \\
22$-$11$-$2016   &    E       & ~0,~4,0 &	17$-$09$-$2017   &    E       & ~0,~4,0 &	28$-$12$-$2018   &    A       & 34,34,1 \\
26$-$11$-$2016   &    E       & ~0,~4,0 &	17$-$09$-$2017   &    G       & ~1,~0,0 &	29$-$12$-$2018   &    A       & 33,33,1 \\
06$-$12$-$2016   &    E       & ~0,~5,0 &	28$-$09$-$2017   &    G       & ~1,~0,0 &	29$-$12$-$2018   &    G       & ~1,~0,0 \\
08$-$12$-$2016   &    E       & ~0,~4,0 &	29$-$09$-$2017   &    G       & ~1,~0,0 &	25$-$01$-$2019   &    G       & ~1,~0,0 \\
10$-$12$-$2016   &    G       & ~1,~0,0 &	12$-$10$-$2017   &    G       & ~1,~0,0 &	16$-$02$-$2019   &    G       & ~1,~0,0 \\
21$-$12$-$2016   &    E       & ~0,~6,0 &       14$-$10$-$2017   &    E       & ~0,~4,0 &	22$-$02$-$2019   &    G       & ~1,~0,0 \\
30$-$12$-$2016   &    A       & 37,37,1 &	22$-$10$-$2017   &    G       & ~1,~0,0 &	28$-$02$-$2019   &    D       & 94,94,1 \\
03$-$01$-$2017   &    E       & ~0,~5,0 &	24$-$10$-$2017   &    E       & ~0,~4,0 &	06$-$03$-$2019   &    D       & 84,84,0 \\
06$-$01$-$2017   &    E       & ~0,~4,0 &	28$-$10$-$2017   &    G       & ~1,~0,0 &	09$-$03$-$2019   &    D       & 17,17,0 \\
13$-$01$-$2017   &    G       & ~1,~0,0 &	11$-$11$-$2017   &    E       & ~0,~4,0 &	09$-$03$-$2019   &    G       & ~1,~0,0 \\
14$-$01$-$2017   &    G       & ~1,~0,0 &	19$-$11$-$2017   &    G       & ~1,~0,0 &	12$-$03$-$2019   &    D       & 67,67,1 \\
17$-$01$-$2017   &    E       & ~0,~4,0 &	21$-$11$-$2017   &    E       & ~0,~6,0 &	15$-$03$-$2019   &    G       & ~1,~0,0 \\
19$-$01$-$2017   &    A       & 27,27,1 &	26$-$11$-$2017   &    E       & ~0,~4,0 &	28$-$03$-$2019   &    F       & 61,61,0 \\
02$-$02$-$2017   &    E       & ~0,~6,0 &	09$-$12$-$2017   &    E       & ~0,~4,0 &	29$-$03$-$2019   &    D       & 22,22,0 \\
13$-$02$-$2017   &    E       & ~0,~4,0 &	16$-$12$-$2017   &    G       & ~1,~0,0 &	30$-$03$-$2019   &    D       & 138,138,0 \\
17$-$02$-$2017   &    G       & ~1,~0,0 &	19$-$12$-$2017   &    E       & ~0,~4,0 &	06$-$04$-$2019   &    D       & 58,58,1 \\
18$-$02$-$2017   &    G       & ~1,~0,0 &	02$-$01$-$2018   &    E       & ~0,~6,0 &	07$-$04$-$2019   &    D       & ~1,~1,0 \\
19$-$02$-$2017   &    E       & ~0,~4,0 &	10$-$01$-$2018   &    B       & 20,20,1 & 	12$-$04$-$2019   &    G       & ~1,~0,0 \\
\hline 
\end{tabular}}
\end{table*}

The key motivation of this paper is to study the multiband optical flux and spectral variability properties of the TeV blazar 1ES 0806$+$524 on diverse timescales. We present the first extensive multiband optical photometric observations of the blazar 1ES 0806$+$524 in $V$, $R$, and $I$ bands from 2011 January 25 to 2019 April 12 using seven optical telescopes. We investigated the flux and spectral variability properties on intraday and longer timescales. We also extracted optical SEDs of the blazar during the observing period.

The paper is organized as follows: Section \ref{sec:data} gives details of observation and data reduction procedures; Section \ref{sec:analysis} describes the analysis techniques we used. Results of our variability analysis are given in Section \ref{sec:res}, while Section \ref{sec:diss} presents a discussion. Finally, a summary of our results is given in Section \ref{sec:summ}.

\begin{figure*}
\centering
\includegraphics[width=18cm, height=18cm]{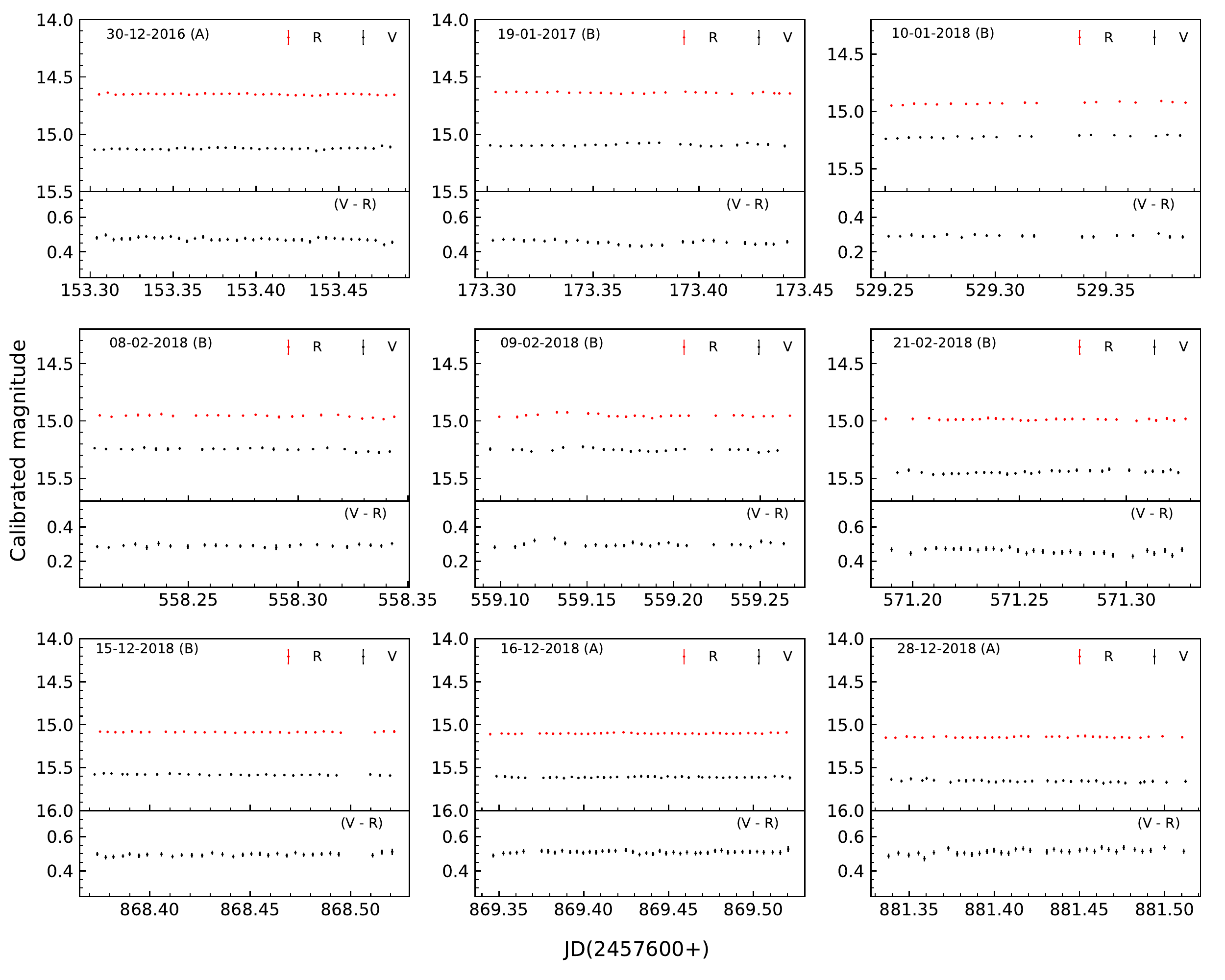}
\caption{\label{fig:lc}Optical IDV LCs of the TeV HBL 1ES 0806$+$524. The observation date and the telescope code are given in each plot. Top panel: $V$ and $R$ band LCs. Bottom panel: Variation of $V-R$ color with time. }
\end{figure*}
\addtocounter{figure}{-1}
\begin{figure*}
\centering
\includegraphics[width=18cm, height=18cm]{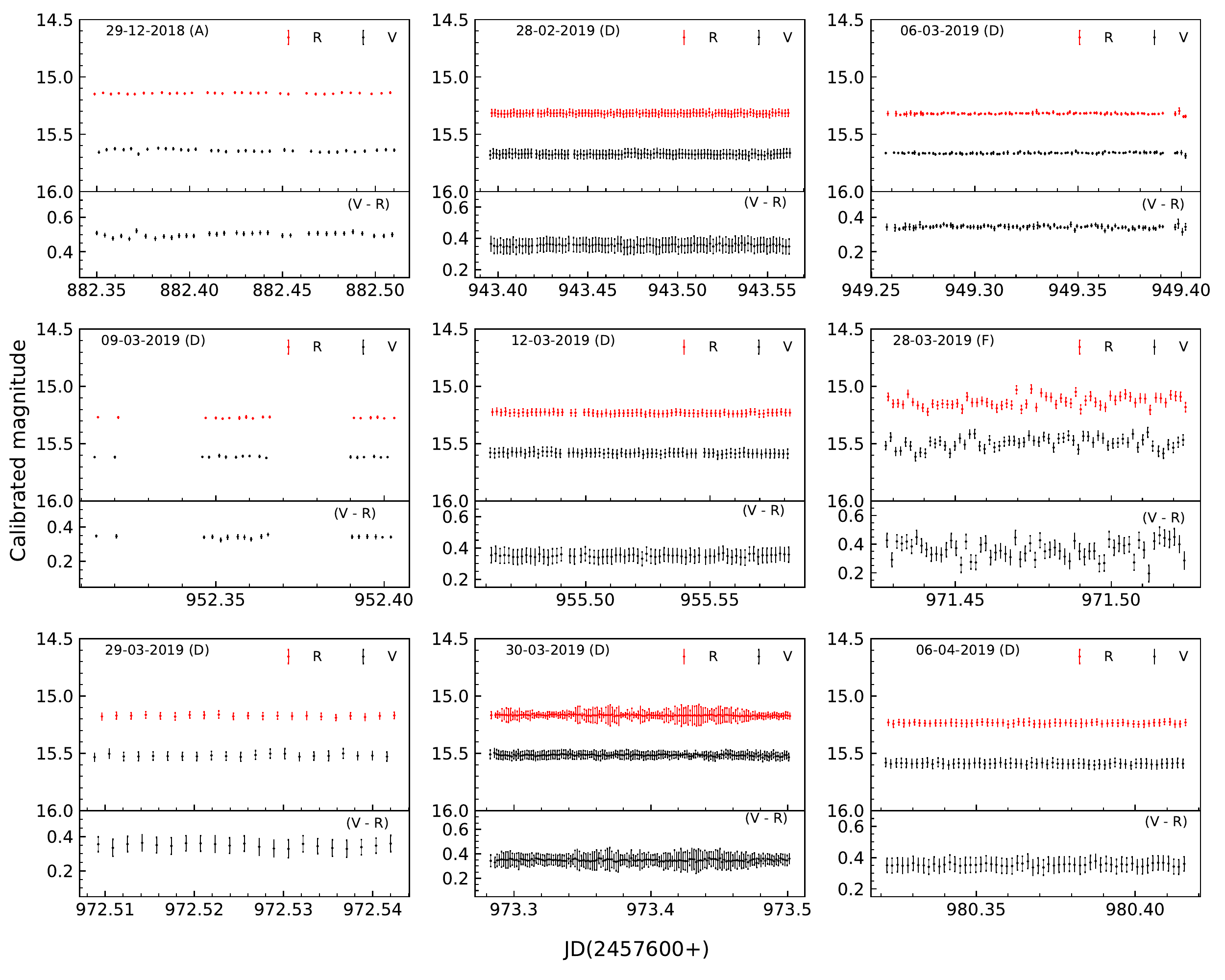}
\caption{\label{fig:lc1}(Continued.) }
\end{figure*}

\section{Observations and Data Reduction} \label{sec:data}

We performed optical photometric observations of the TeV blazar 1ES 0806$+$524 from 2011 January 25 to 2019 April 12. Our observations were carried out using a total of seven ground-based telescopes; two telescopes in India, two in Bulgaria, one in Georgia, one in Serbia, and one in Spain. We observed the blazar for 153 nights during which a total of 2263 image frames were collected in $V$, $R$, and $I$ optical bands. 
The technical details of the telescopes and the instruments used for observations are given in Table \ref{tab:teles}. The complete observation log of the optical photometric observations of the TeV HBL 1ES 0806$+$524 is given in Table \ref{tab:obs_log}.

We monitored the blazar 1ES 0806$+$524 for a total of 10 nights between 2016 December 30 and 2018 December 29 with two Ritchey$-$Chretien (RC) Cassegrain reflector telescopes in India: 1.3 m ($f/4$) Devasthal Fast Optical Telescope (DFOT) and 1.04 m ($f/13$) Sampuranand Telescope (ST). Both of these telescopes are equipped with Johnson-Cousins {\it UBVRI} filters. Using these telescopes, we observed the blazar quasi-simultaneously in $V$ and $R$ optical filters for $\sim 3-4$ hr during each observing night. Single $I$ band image frames were also taken on each night. The cleaning (bias-subtraction, flat-fielding and cosmic-ray removal) of the raw CCD images was done using IRAF\footnote{Image Reduction and Analysis Facility (IRAF) is distributed by the National Optical Astronomy Observatory, which is operated by the Association of Universities for Research in Astronomy (AURA) under a cooperative agreement with the National Science Foundation.}. We then performed the aperture photometry in each cleaned CCD image using DAOPHOT\footnote{Dominion Astronomical Observatory Photometry} \RNum{2} software to get the instrumental magnitudes of the blazar and the stars in the image frame. Details of the data reduction process are given in \cite{2019ApJ...871..192P}. 

Observations with the 60 cm Cassegrain telescope located at the Astronomical Observatory (AO) Belogradchik, Bulgaria, were carried out in optical $V$, $R$ and $I$ bands for a total of 13 nights between 2011 January 25 and 2015 December 14. We also observed the source quasi-simultaneously in $V$ and $R$ bands on 2019 March 28 with the 50/70 cm Schmidt telescope at the National Astronomical Observatory (NAO), Rozhen, Bulgaria. The data reduction process for these telescopes is explained in detail in \cite{2012MNRAS.420.3147G}. 

Optical photometric observations of 1ES 0806$+$524 were conducted in the $R$-band from 2016 January 18 to 2018 April 23 for a total of 66 nights using the 70 cm ($f/3$) Meniscus telescope located at Abastumani Observatory, Georgia. The telescope is equipped with an Apogee CCD camera AP6E and Cousins $R$ filter. Cleaning of the raw data was done using IRAF and the photometry was performed using DAOPHOT \RNum{2}.

Quasi-simultaneous monitoring of the source in $V$ and $R$ filters was also performed using the 1.4 m telescope located at Astronomical Station Vidojevica (ASV), Serbia. The telescope has a CCD camera Andor iKon-L and is equipped with Johnson-Cousins UBVRI broadband filters. With this telescope, we observed the blazar for a total of nine nights between 2018 December 2 and 2019 April 7. The optical photometric data analysis procedure we followed is given in \cite{2019AJ....157...95G}.

We also carried out optical $V$-band observations of the TeV blazar 1ES 0806$+$524 on 54 nights between 2016 August 19 and 2019 April 12 using a Celestron C14 XLT 35.6 cm telescope at Las Casqueras, Spain. The CCD raw images were cleaned using the standard process and photometry was done using MaxIm DL software. 

The instrumental magnitudes of the blazar were calibrated using the standard star C4, taken from the blazar finding chart\footnote{\url{https://www.lsw.uni-heidelberg.de/projects/extragalactic/charts/0806+526.html}}, having brightness close to that of the blazar.

We monitored the TeV HBL 1ES 0806$+$524 for a total of 153 nights in optical $V$, $R$, and $I$ bands during our campaign. Out of these 153 nights, 85 had multiple observations made of the source in a specific filter. However, to search for IDV flux and spectral variations, we only selected nights in which at least 15 measurements were taken in an optical band by a telescope. By applying this criterion, 18 nights qualified for the IDV flux and spectral variability analysis; those nights with more than 15 observations had data taken in the $V$ and $R$ bands.

\section{Analysis Techniques} \label{sec:analysis}
To search for intraday variations in the optical light curves (LCs) of 1ES 0806$+$524, we have used two powerful statistical tests {that have frequently been employed recently to examine AGN variability}; the power-enhanced {\it F}-test and the nested analysis of variance (ANOVA) test \citep{2014AJ....148...93D, 2015AJ....150...44D}. In both the tests, multiple field stars are involved as comparison stars in the analysis to produce a more reliable result.  A large enough number of independent observations that allow us to properly look for IDV were made on 18 nights between 30 December 2016 and 6 April 2019.

\subsection{Power-enhanced {\it F}-test}\label{sec:f_test}
In the power-enhanced {\it F}-test, the brightest unsaturated star is used as a reference star to get the differential light curves (DLCs) of the blazar and the other comparison stars \citep{2015MNRAS.452.4263G, 2016MNRAS.460.3950P, 2017MNRAS.466.2679K}. This statistical test is explained in detail in \cite{2019ApJ...871..192P}. First, we estimated the variance, $s_{blz}^2$, of the blazar DLCs. We then calculated the combined variance, $s_c^2$, of the DLCs of all the comparison stars using Equation (2) of \cite{2019ApJ...871..192P}. The value of power-enhanced {\it F}-statistics, $F_{enh}$, is calculated as:
\begin{equation} 
F_{enh} = \frac{s_{blz}^2}{s_c^2},
\end{equation}   
In this work, we have always observed three or more stars that are close in magnitude to the blazar and are in its field. Out of these nearby stars, we selected the brightest star as the reference star and the remaining ($k$) field stars as the comparison stars. Since the blazar and all the comparison stars are in the same field, they have the same number of observations ($N$). We estimated the critical value ($F_c$) of $F-$statistics at $\alpha = 0.01$ (i.e., at nominal 99\% confidence level) using the number of degrees of freedom in the numerator, $\nu_{blz} = N-1$ and in denominator, $\nu_c = k(N-1)$. 
We then compared the $F_{enh}$ value with the critical value ($F_c$). A light curve is considered variable (V) if $F_{enh} \geq F_c$; otherwise, we label it nonvariable (NV). \\
\\
\subsection{Nested {\it ANOVA}}\label{sec:anova}
In the nested ANOVA test, all the comparison stars are used as reference stars to get the DLCs of the AGN \citep{2015AJ....150...44D,2019ApJ...871..192P}. These different DLCs of a blazar are then grouped such as each group has five points. The $F$ statistic is calculated as $F = MS_{G}/MS_{O(G)}$, where $MS_G$ and $MS_{O(G)}$, calculated using Equation (4) of \cite{2015AJ....150...44D}, are the mean square due to groups and the mean square due to nested observations in groups, respectively. A light curve is called variable (V) if the $F-$statistic $\geq$ $(F_c)$ at $\alpha$ = 0.01, otherwise nonvariable (NV). 

The results of both the statistical tests are given in Table \ref{tab:var_res}, where a light curve is conservatively labeled as variable (V) only if both the tests found significant variations in it, otherwise it is labeled as nonvariable (NV), though of course there may be weak intrinsic variability even in some of those cases.  \\
\\
\subsection{Intraday Variability Amplitude}
We calculated the amplitude ($A$, in percent) of flux and color variations on both IDV and LTV timescales using the following equation \citep{1996A&A...305...42H}:
\begin{equation}\label{eq:var_amp}
A = 100\times \sqrt{(A_{max}-A_{min})^2 - 2 \sigma^2} ,
\end{equation}
where $A_{max}$ and $A_{min}$ denote the maximum and minimum calibrated magnitudes, respectively, in the light curve of blazar, and $\sigma$ represents the mean error. 

\section{Results}\label{sec:res}

\subsection{Variability on IDV Timescale} \label{sec:IDV}
\subsubsection{Flux Variability} \label{sec:idv_flux}
The calibrated optical IDV LCs of the TeV blazar 1ES 0806$+$524 in $V$ and $R$ bands are plotted in the upper panel of each plot in Figure \ref{fig:lc}. As can been seen from the plots in Figure \ref{fig:lc}, all the IDV LCs are either flat or exhibit quite small fluctuations. 

We examined the $V$ and $R$ band LCs of 1ES 0806$+$524 on the 18 nights with dense sampling for intraday variations using the two very reliable tests discussed above: the power-enhanced {\it F}-test and the nested ANOVA test. The results of the analysis are given in Table \ref{tab:var_res}. We found no IDV that was statistically significant by both tests on any night except on 2018 January 10. The amplitude of variations in the $V$ and $R$ band LCs on 2018 January 10, calculated using Equation \ref{eq:var_amp}, are only 3.5 \% and 3.9 \%, respectively. There were 11 more occasions when the source showed IDV by the nested-ANOVA test but does not show IDV by the power enhanced F-test, so we did not consider those nights as definitely possessing IDV. 

\subsubsection{Spectral Variability} \label{sec:idv_spec}
To study spectral variability of the blazar 1ES 0806+524 on IDV timescales, the $V-R$ color indices (CIs) were calculated for each pair of $V$ and $R$ magnitudes and plotted against time (color-time), in the bottom panels of Figure \ref{fig:lc}. 

We then examined color-time plots using the two statistical tests, discussed in section \ref{sec:f_test} and section \ref{sec:anova}. The results of the analysis are also shown in Table \ref{tab:var_res}. We found no significant temporal variations in $V-R$ CIs on IDV timescales. 

\begin{table*}
\caption{Results of IDV analysis of 1ES 0806$+$524} 
\label{tab:var_res} 
\centering 
\resizebox{0.8\textwidth} {!}{ 
\begin{tabular}{lcccccccccc} 
\hline\hline 
Observation date & Band & \multicolumn{3}{c}{{\it Power-enhanced F-test}} & \multicolumn{3}{c}{{\it Nested ANOVA}} & Status & Amplitude\\
\cmidrule[0.03cm](r){3-5}\cmidrule[0.03cm](r){6-8} dd-mm-yyyy & & DoF($\nu_1$,$\nu_2$ ) & $F_{enh}$ & $F_c$ & DoF($\nu_1$,$\nu_2$ ) & $F$ & $F_c$ & &$\%$ \\
\hline
30-12-2016 & V & 36,108 & 0.66 & 1.81 & 6, 28 & 3.83 & 3.53 & NV & - \\ 
& R & 36,108 & 0.50 & 1.81 & 6, 28 & 2.21 & 3.53 & NV & - \\ 
& V-R & 36,108 & 0.61 & 1.81 & 6, 28 & 1.96 & 3.53 & NV & - \\ 

19-01-2017 & V & 26, 78 & 0.47 & 2.00 & 4, 20 & 1.91 & 4.43 & NV & - \\ 
& R & 26, 78 & 0.41 & 2.00 & 4, 20 & 1.32 & 4.43 & NV & - \\ 
& V-R & 26, 78 & 0.80 & 2.00 & 4, 20 & 2.44 & 4.43 & NV & - \\ 

10-01-2018 & V & 19, 38 & 2.81 & 2.42 & 3, 16 & 14.27 & 5.29 & V & 3.46 \\ 
& R & 19, 38 & 3.08 & 2.42 & 3, 16 & 27.56 & 5.29 & V & 3.88 \\ 
& V-R & 19, 38 & 0.44 & 2.42 & 3, 16 & ~0.37 & 5.29 & NV & - \\ 

08-02-2018 & V & 24, 48 & 0.85 & 2.20 & 4, 20 & 14.89 & 4.43 & NV & - \\ 
& R & 24, 48 & 0.70 & 2.20 & 4, 20 & 13.09 & 4.43 & NV & - \\ 
& V-R & 24, 48 & 0.67 & 2.20 & 4, 20 & ~1.10 & 4.43 & NV & - \\ 

09-02-2018 & V & 24, 48 & 0.84 & 2.20 & 4, 20 & 3.11 & 4.43 & NV & - \\ 
& R & 24, 48 & 0.71 & 2.20 & 4, 20 & 0.98 & 4.43 & NV & - \\ 
& V-R & 24, 48 & 0.96 & 2.20 & 4, 20 & 0.35 & 4.43 & NV & - \\ 

21-02-2018 & V & 29, 87 & 0.59 & 1.94 & 5, 24 & 1.58 & 3.90 & NV & - \\ 
& R & 29, 87 & 0.62 & 1.94 & 5, 24 & 1.99 & 3.90 & NV & - \\ 
& V-R & 29, 87 & 1.33 & 1.94 & 5, 24 & 1.49 & 3.90 & NV & - \\ 

15-12-2018 & V & 29, 87 & 0.57 & 1.94 & 5, 24 & 3.00 & 3.90 & NV & - \\ 
& R & 29, 87 & 0.54 & 1.94 & 5, 24 & 4.97 & 3.90 & NV & - \\ 
& V-R & 29, 87 & 1.31 & 1.94 & 5, 24 & 0.55 & 3.90 & NV & - \\ 

16-12-2018 & V & 40,120 & 0.56 & 1.76 & 7, 32 & 1.98 & 3.26 & NV & - \\ 
& R & 40,120 & 0.49 & 1.76 & 7, 32 & 2.44 & 3.26 & NV & - \\ 
& V-R & 40,120 & 0.88 & 1.76 & 7, 32 & 0.58 & 3.26 & NV & - \\ 

28-12-2018 & V & 33, 99 & 1.20 & 1.86 & 5, 24 & 5.11 & 3.90 & NV & - \\ 
& R & 33, 99 & 1.04 & 1.86 & 5, 24 & 7.23 & 3.90 & NV & - \\ 
& V-R & 33, 99 & 0.88 & 1.86 & 5, 24 & 2.27 & 3.90 & NV & - \\ 

29-12-2018 & V & 32, 96 & 0.77 & 1.88 & 5, 24 & 0.99 & 3.90 & NV & - \\ 
& R & 32, 96 & 0.22 & 1.88 & 5, 24 & 0.49 & 3.90 & NV & - \\ 
& V-R & 32, 96 & 0.68 & 1.88 & 5, 24 & 1.27 & 3.90 & NV & - \\ 

28-02-2019  & V &  93,279 & 1.96 & 1.46 &  17, 72 & 2.14 & 2.23 & NV &   - \\ 
& R &  93,279 & 1.75 & 1.46 &  17, 72 & 0.55 & 2.23 & NV &  - \\ 
& V-R &  90,270 & 1.52 & 1.47 &  17, 72 & 1.64 & 2.23 & NV &  - \\

06-03-2019 & V & 83,166 & 1.07 & 1.54 & 15, 64 & 2.46 & 2.33 & NV & - \\ 
& R & 83,166 & 0.80 & 1.54 & 15, 64 & 3.63 & 2.33 & NV & - \\ 
& V-R & 83,166 & 0.79 & 1.54 & 15, 64 & 2.28 & 2.33 & NV & - \\

09-03-2019 & V & 16, 32 & 1.37 & 2.62 & 2, 12 & 0.80 & 6.93 & NV & - \\ 
& R & 16, 32 & 0.97 & 2.62 & 2, 12 & 0.08 & 6.93 & NV & - \\ 
& V-R & 16, 32 & 0.92 & 2.62 & 2, 12 & 0.15 & 6.93 & NV & - \\ 

12-03-2019  & V &  66,198 & 1.51 & 1.56 &  12, 52 & 3.18 & 2.55 & NV &   - \\ 
& R &  66,198 & 1.26 & 1.56 &  12, 52 & 5.55 & 2.55 & NV &   - \\ 
& V-R &  64,192 & 1.18 & 1.57 &  12, 52 & 0.98 & 2.55 & NV &  - \\ 

28-03-2019 & V & 60,180 & 0.95 & 1.60 & 11, 48 & 2.06 & 2.64 & NV & - \\ 
& R & 60,180 & 0.77 & 1.60 & 11, 48 & 0.65 & 2.64 & NV & - \\ 
& V-R & 60,180 & 0.85 & 1.60 & 11, 48 & 1.53 & 2.64 & NV & - \\ 

29-03-2019 & V & 21, 63 & 1.02 & 2.16 & 3, 16 & 1.94 & 5.29 & NV & - \\ 
& R & 21, 63 & 0.76 & 2.16 & 3, 16 & 2.34 & 5.29 & NV & - \\ 
& V-R & 21, 63 & 0.80 & 2.16 & 3, 16 & 6.50 & 5.29 & NV & - \\ 

30-03-2019 & V & 137,274 & 1.02 & 1.40 & 26,108 & 0.96 & 1.93 & NV & - \\ 
& R & 137,274 & 0.99 & 1.40 & 26,108 & 0.60 & 1.93 & NV & - \\ 
& V-R & 137,274 & 0.99 & 1.40 & 26,108 & 0.51 & 1.93 & NV & - \\ 

06-04-2019  & V &  57,171 & 1.65 & 1.61 &  10, 44 & 1.13 & 2.75 & NV &  - \\ 
& R &  57,171 & 0.97 & 1.61 &  10, 44 & 2.23 & 2.75 & NV &  - \\ 
& V-R &  55,165 & 0.94 & 1.63 &  10, 44 & 1.90 & 2.75 & NV &  - \\ 
\hline 
\end{tabular}}
\end{table*}

\subsection{Variability on LTV Timescale} \label{sec:LTV}

\subsubsection{Flux Variability} \label{sec:ltv_flux}
\begin{figure*}
\centering
\includegraphics[width=15cm, height=9cm]{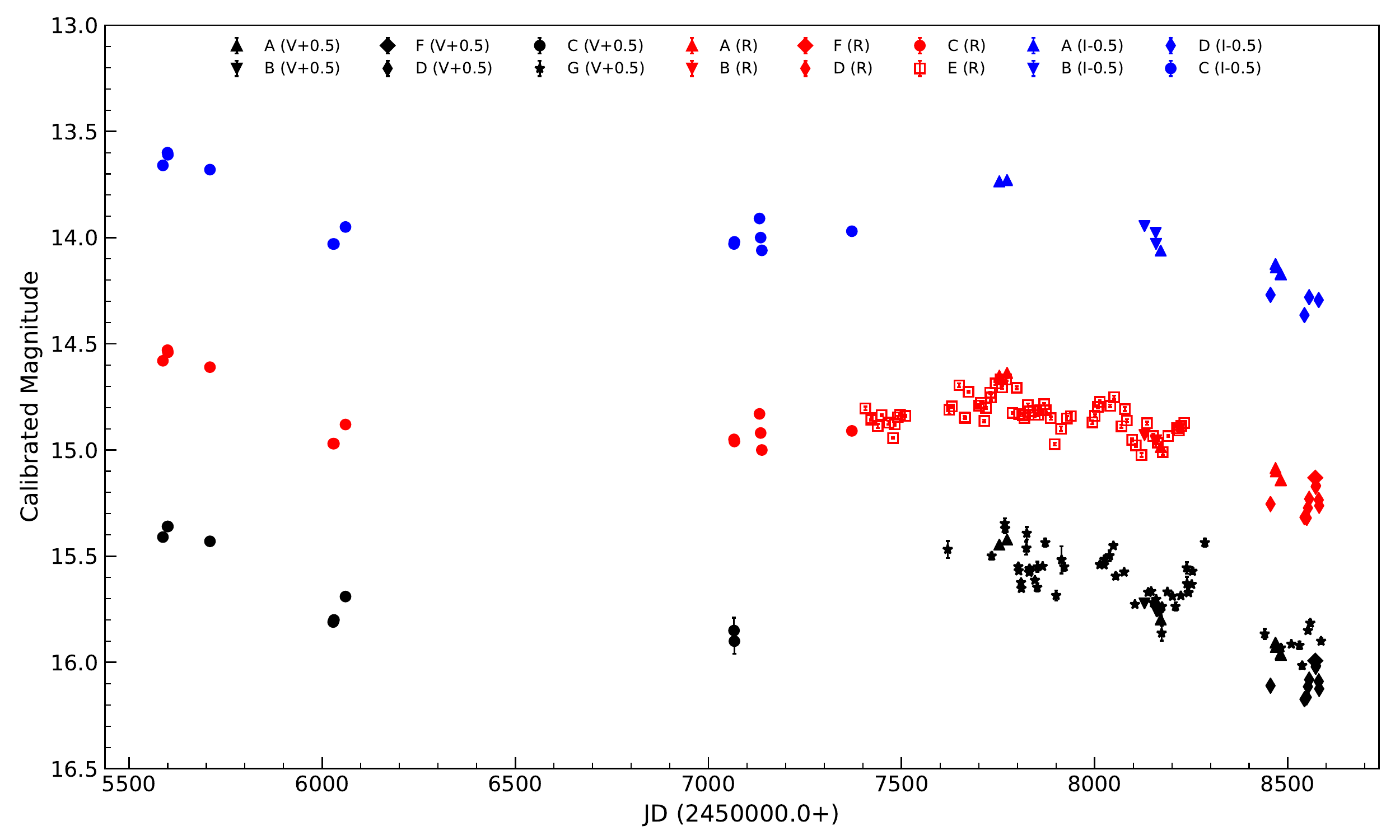} 
\caption{LTV optical (VRI) light curves of 1ES 0806$+$524; they are shown in black ($V$); red ($R$) and blue ($I$), respectively.} 
\label{fig:ltv}
\end{figure*}

The LTV calibrated light curves of the blazar 1ES 0806$+$524 in $V$, $R$ and $I$ bands for the entire observing period are plotted in Figure \ref{fig:ltv}, where the daily averaged calibrated magnitudes are plotted against time. We have shifted the $V$ and $I$ band LCs by $+0.5$ and $-0.5$ mag, respectively, to make the long term optical variability patterns more easily visible. Variations on LTV timescales can be clearly seen in all three optical LCs. The amplitudes of variation in $V$, $R$, and $I$ optical wavebands are 84.3 \%, 81.6 \%, and 76.5 \%, respectively. The average magnitudes in $V$, $R$ and $I$ bands were 15.45, 15.06, and 14.51, respectively. During our eight year long observing campaign, the blazar 1ES 0806$+$524 was detected in the brightest state of $R_{mag}=$14.53 on 6 February 2011, while the faintest magnitude observed was $R_{mag}=$15.35 on 6 March 2019. Our brightest and faintest magnitudes of 1ES 0806$+$524 are very close to  previous observations of this blazar.  \cite{2015MNRAS.451..739A} observed the blazar in the brightest state of $R_{mag}=$14.38 and \cite{2012JPhCS.355a2013R} reported the faintest state of $R_{mag}=$15.27.

\subsubsection{Spectral Variability} \label{sec:ltv_spec}
\begin{figure}
\centering
\includegraphics[width=8.5cm, height=8.5cm]{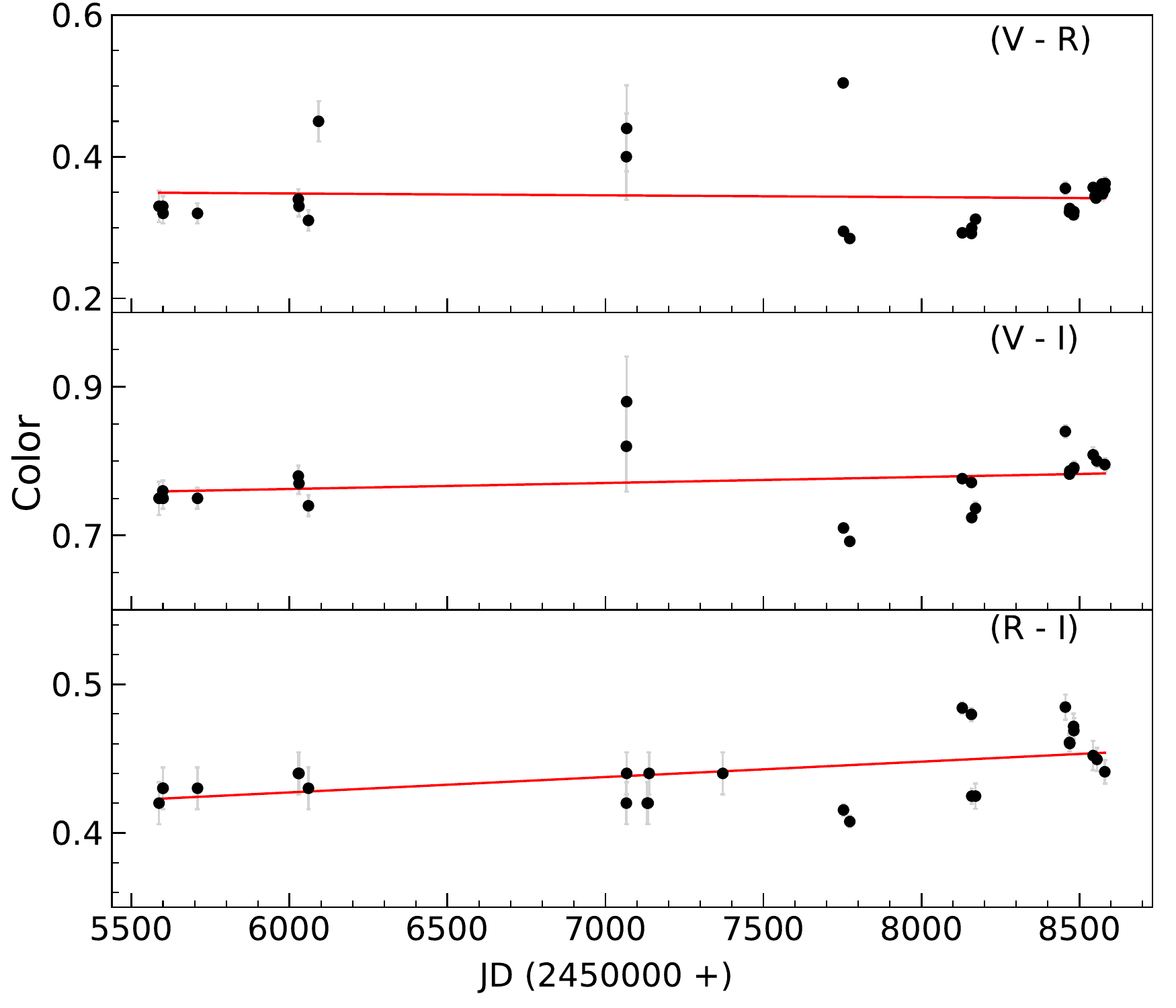}
\caption{Optical color variations of 1ES 0806$+$524 with time during the entire monitoring period.} 
\label{fig:ltv_ct}
\end{figure}

\begin{figure}
\centering
\includegraphics[width=8.5cm, height=8.5cm]{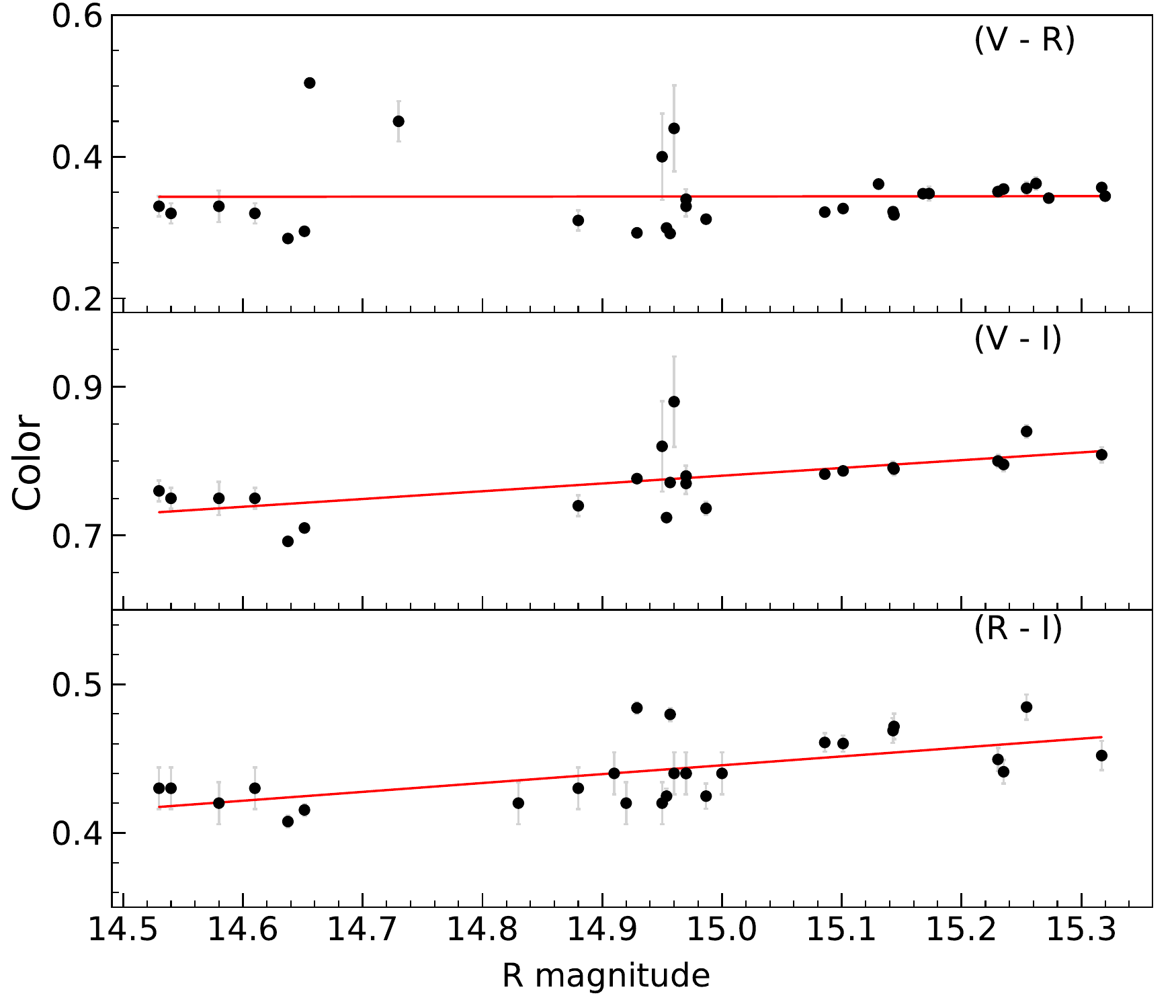}
\caption{Optical color variations of 1ES 0806$+$524 against R magnitude.} 
\label{fig:ltv_cm}
\end{figure}

To investigate possible spectral changes over longer timescales, we plotted optical color indices, both with respect to time and with respect to R-band magnitude, which has a relatively large overlap with other bands, in Figures \ref{fig:ltv_ct} and \ref{fig:ltv_cm}, respectively. Visual inspection of both the figures shows clear color variations. However, to check whether these variations are systematic or not, we fitted each panel in Figures \ref{fig:ltv_ct} and \ref{fig:ltv_cm} with a straight line of the form $Y = mX + c$. The results of the fit for color-time and color-magnitude plots are given in Tables \ref{tab:ltv_ct} and \ref{tab:ltv_cm}, respectively. Only the $R-I$ color shows a significant systematic variations with time, while systematic color variations with respect to $R$ magnitude were found in $V-I$ and $R-I$ colors.

\subsection{Spectral Energy Distribution} \label{sec:sed}
\begin{figure*}
\centering
\includegraphics[width=18cm, height=19cm]{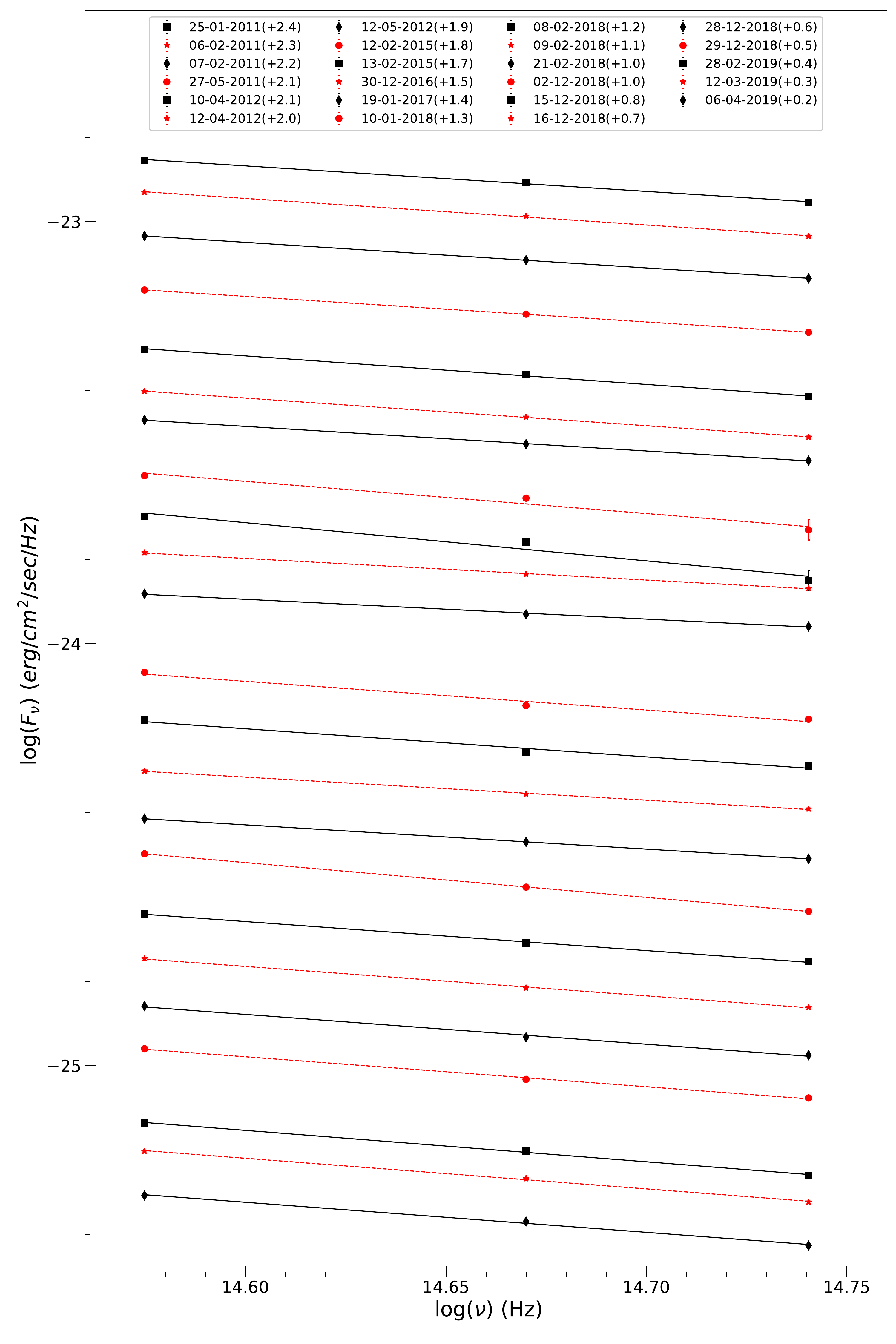}
\caption{SEDs of 1ES 0806$+$524 in $V$, $R$, and $I$ bands.} 
\label{fig:sed}
\end{figure*}

\begin{figure}
\centering
\includegraphics[width=8.5cm, height=6cm]{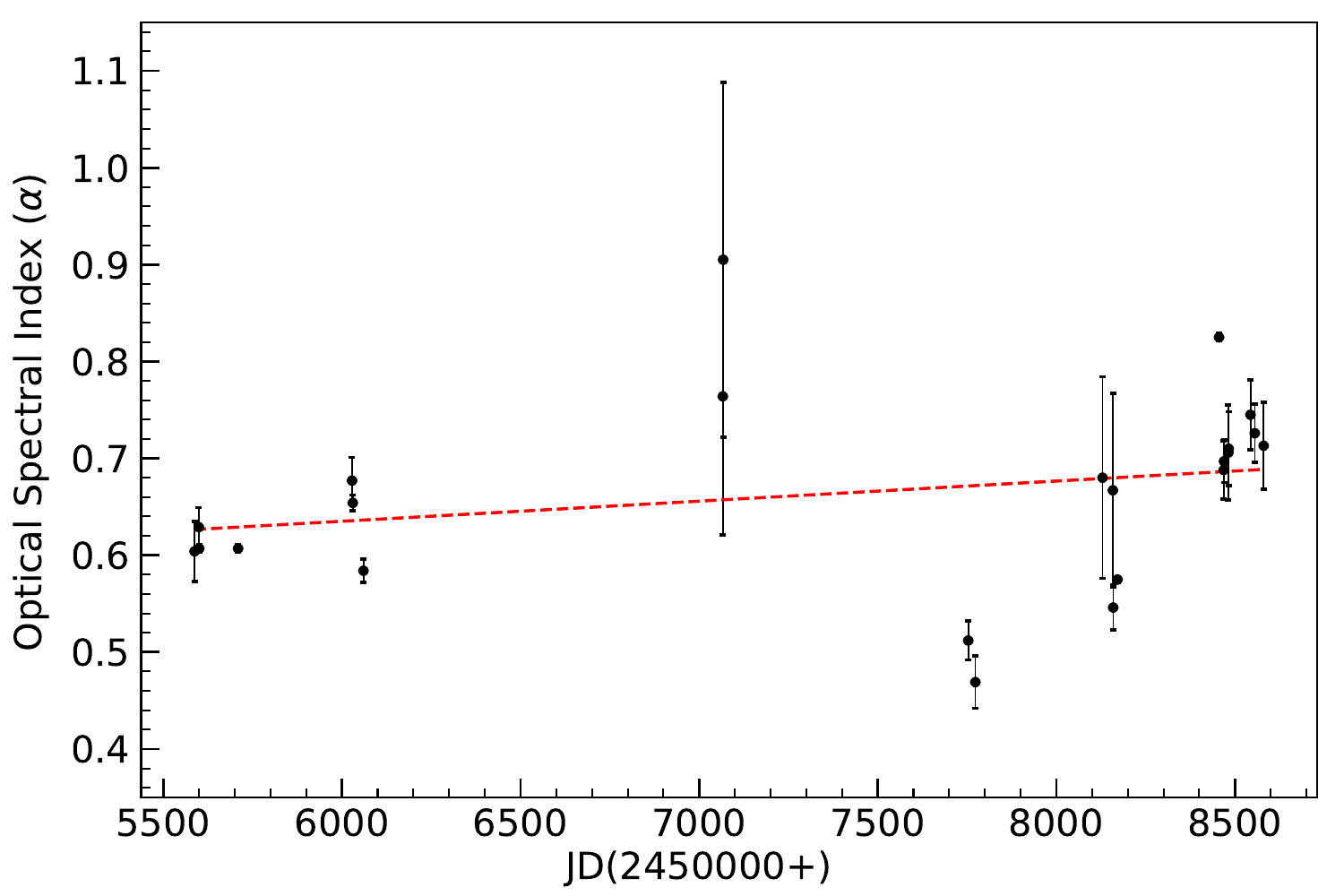}
\includegraphics[width=8.5cm, height=6cm]{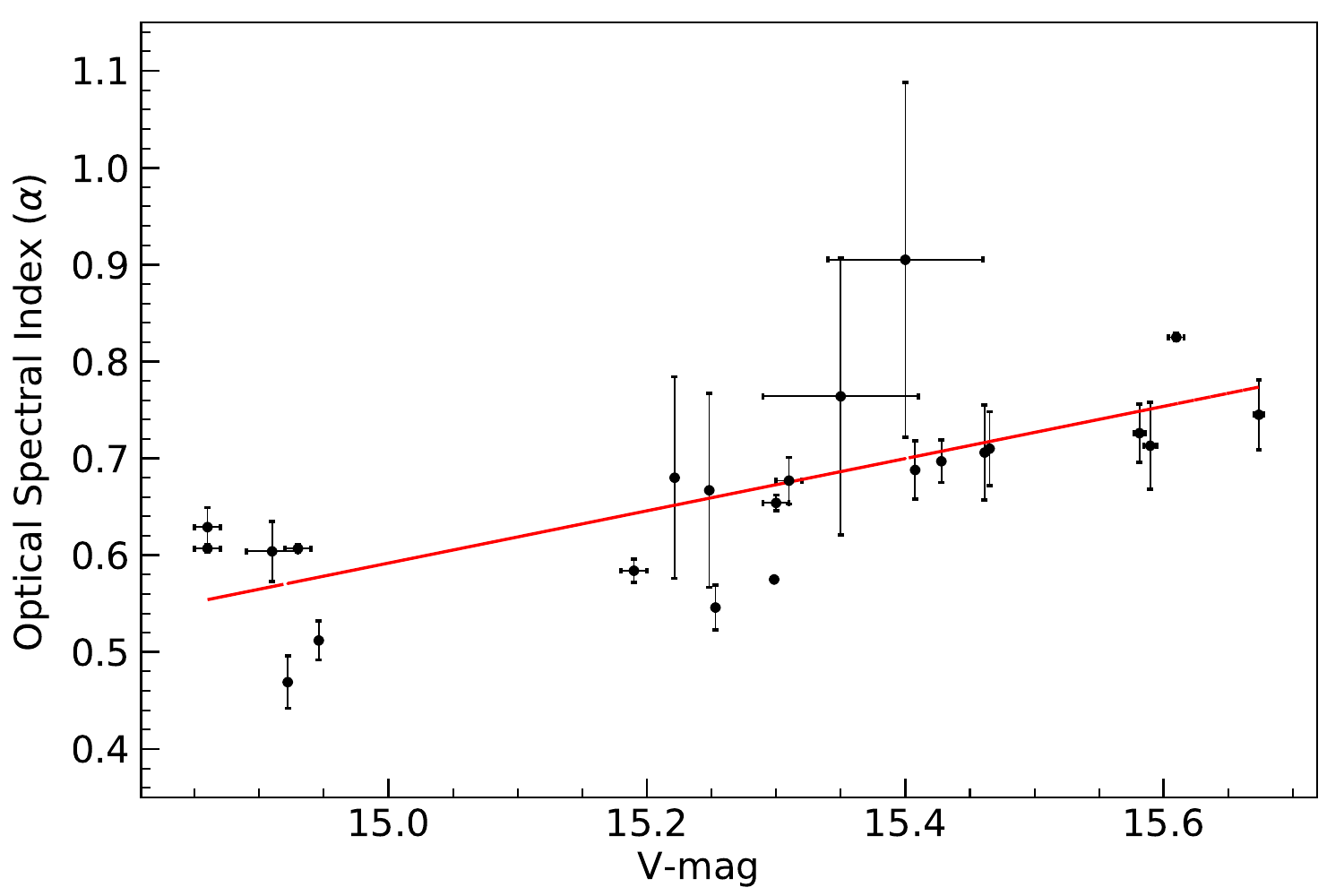}
\caption{Variation of optical spectral index with respect to time (top) and V-magnitude (bottom).} 
\label{fig:alpha}
\end{figure}

For a detailed study of spectral variations on longer timescales, we extracted the optical (VRI) SEDs of 1ES 0806+524 for those nights in which observations were taken in all three optical bands. For this, we first subtracted the galactic extinction, $A_{\lambda}$, taken from 
the NASA Extragalactic Database (NED\footnote{\url{https://ned.ipac.caltech.edu/}}) from the calibrated magnitudes and then converted them into extinction corrected flux densities, $F_{\nu}$. The optical SEDs of 1ES 0806+524, in $log(F_{\nu}) - log(\nu)$ representation, are shown in Figure \ref{fig:sed}. Since the optical continuum spectra of blazars are usually well represented by a simple power law ($F_{\nu} \propto \nu^{-\alpha}$, $\alpha$ being the optical spectral index), we fitted each SED with a straight line ($log(F_{\nu}) = -\alpha~log(\nu)+ C$) to obtain the optical spectral indices of the blazar. The results of linear fits are given in Table \ref{tab:sed}. 
The derived spectral indices range from 0.469$\pm$0.027 to 0.905$\pm$0.183, quantifying the obvious spectral variation in the optical band. However, the errors of some of the spectral indices are very large. The weighted mean optical spectral index of 1ES 0806$+$524 during our observation campaign is 0.639$\pm$0.002, which is similar to those found by \cite{2012MNRAS.425.3002G}. The behavior of optical spectral indices is often studied with respect to time and with respect to $V-$magnitude \citep[e.g.][and references therein]{2019AJ....157...95G}. So, we plotted the spectral indices with respect to time and with respect to $V-$band magnitude in the top and bottom panels of Figure \ref{fig:alpha}, respectively. To search for any systematic variations in the spectral index, we fitted each panel in Figure \ref{fig:alpha} with a straight line. The results of the fits are given in Table \ref{tab:alpha_tv}. No systematic temporal variation is found in the optical spectral index. The spectral index shows a significant positive correlation with $V-$band magnitude, indicating a BWB behavior.  We also investigated the variation of spectral indices with respect to R-magnitude and with respect to I-magnitude and found similar trends. 

\begin{table}

\caption{Variation of color indices with respect to time on LTV timescales.} 
\label{tab:ltv_ct} 
\centering 
\resizebox{0.5\textwidth} {!}{ 
\begin{tabular}{lcccc} 
\hline\hline 
Color Index & $m_1^a$ & $c_1^a$ & $r_1^a$ & $p_1^a$ \\
& & & & \\ 
\hline 
V - R &$-2.592e-06 \pm 7.807e-06 $ &  0.364 & -0.062 & 7.423e-01 \\
V - I &$8.064e-06 \pm 7.542e-06 $ &  0.714 &  0.227 & 2.971e-01 \\
R - I &$1.035e-05 \pm 3.438e-06 $ &  0.365 &  0.516 & 5.892e-03 \\
\hline 
\end{tabular}}\\
$^am_1$ = slope and $c_1$ = intercept of CI against time; $r_1$ = Correlation coefficient; $p_1$ = null hypothesis probability
\end{table}

\begin{table}
\caption{Variation of color indices with respect to R magnitude on LTV timescales.} 
\label{tab:ltv_cm} 
\centering 
\resizebox{0.5\textwidth} {!}{ 
\begin{tabular}{lcccc} 
\hline\hline 
Color Index & $m_2^a$ & $c_2^a$ & $r_2^a$ & $p_2^a$ \\
& & & & \\ 
\hline 
V - R &$1.261e-03 \pm 3.580e-02 $ &  0.325 &  0.007 & 9.721e-01 \\
V - I &$1.045e-01 \pm 3.008e-02 $ & -0.787 &  0.604 & 2.276e-03 \\
R - I &$5.965e-02 \pm 1.543e-02 $ & -0.449 &  0.612 & 6.999e-04 \\
\hline 
\end{tabular}}\\
$^am_2$ = slope and $c_2$ = intercept of CI against R magnitude; $r_2$ = Correlation coefficient; $p_2$ = null hypothesis probability
\end{table}

\begin{table}
\caption{Straight line fits to optical SEDs of TeV blazar 1ES 0806$+$524.} 
\label{tab:sed} 
\centering 
\resizebox{0.5\textwidth} {!}{ 
\hskip-1.cm 
\begin{tabular}{lcccc} 
\hline\hline 
Observation date & $\alpha^b$ & $C^b$ & $r_3^b$ & $p_3^b$ \\
dd$-$mm$-$yyyy & & & & \\ 
\hline 
25-01-2011	&$ 0.604 \pm  0.031 $ & -14.047 & -0.999 & 3.308e-02 \\
06-02-2011	&$ 0.629 \pm  0.020 $ & -13.757 & -1.000 & 1.986e-02 \\
07-02-2011	&$ 0.607 \pm  0.004 $ & -14.193 & -1.000 & 3.920e-03 \\
27-05-2011	&$ 0.607 \pm  0.004 $ & -14.321 & -1.000 & 3.920e-03 \\
10-04-2012	&$ 0.677 \pm  0.024 $ & -13.430 & -0.999 & 2.233e-02 \\
12-04-2012	&$ 0.654 \pm  0.008 $ & -13.862 & -1.000 & 7.647e-03 \\
12-05-2012	&$ 0.584 \pm  0.012 $ & -14.962 & -1.000 & 1.327e-02 \\
12-02-2015	&$ 0.764 \pm  0.143 $ & -12.466 & -0.983 & 1.177e-01 \\
13-02-2015	&$ 0.905 \pm  0.183 $ & -10.498 & -0.980 & 1.269e-01 \\
30-12-2016	&$ 0.512 \pm  0.020 $ & -16.323 & -0.999 & 2.454e-02 \\
19-01-2017	&$ 0.469 \pm  0.027 $ & -17.046 & -0.998 & 3.649e-02 \\
10-01-2018	&$ 0.680 \pm  0.104 $ & -14.164 & -0.989 & 9.629e-02 \\
08-02-2018	&$ 0.667 \pm  0.100 $ & -14.466 & -0.989 & 9.495e-02 \\
09-02-2018	&$ 0.546 \pm  0.023 $ & -16.344 & -0.999 & 2.690e-02 \\
21-02-2018	&$ 0.575 \pm  0.003 $ & -16.041 & -1.000 & 3.549e-03 \\
02-12-2018	&$ 0.825 \pm  0.004 $ & -12.480 & -1.000 & 3.086e-03 \\
15-12-2018	&$ 0.688 \pm  0.030 $ & -14.615 & -0.999 & 2.747e-02 \\
16-12-2018	&$ 0.697 \pm  0.022 $ & -14.581 & -1.000 & 1.967e-02 \\
28-12-2018	&$ 0.706 \pm  0.049 $ & -14.572 & -0.998 & 4.450e-02 \\
29-12-2018	&$ 0.710 \pm  0.038 $ & -14.616 & -0.999 & 3.399e-02 \\
28-02-2019	&$ 0.745 \pm  0.036 $ & -14.275 & -0.999 & 3.063e-02 \\
12-03-2019	&$ 0.726 \pm  0.030 $ & -14.624 & -0.999 & 2.602e-02 \\
06-04-2019	&$ 0.713 \pm  0.045 $ & -14.915 & -0.998 & 4.010e-02 \\
\hline 
\end{tabular}}\\
$^b\alpha$ = spectral index and $C$ = intercept of log($F_{\nu}$) against log($\nu$); $r_3$ = Correlation coefficient; $p_3$ = null hypothesis probability
\end{table}

\begin{table}
\caption{Variation of optical spectral index, $\alpha$, with respect to time and V-magnitude during our observing campaign of 1ES 0806+524.} 
\label{tab:alpha_tv} 
\centering 
\resizebox{0.5\textwidth} {!}{ 
\begin{tabular}{lcccc} 
\hline\hline 
Parameter & $m_4^a$ & $c_4^a$ & $r_4^a$ & $p_4^a$ \\
& & & & \\ 
\hline 
$\alpha$ vs time  &$2.075e-05 \pm 1.763e-05$ &  0.511 &  0.249 & 2.523e-01 \\
$\alpha$ vs V-mag &$ 0.270    \pm  0.060$    & -3.454 &  0.700 & 1.994e-04 \\
\hline 
\end{tabular}}\\
$^am_4$ = slope and $c_4$ = intercept of $\alpha$ against time or V magnitude; $r_4$ = Correlation coefficient; $p_4$ = null hypothesis probability
\end{table}

\section{Discussion} \label{sec:diss}
Flux variability on diverse timescales at all electromagnetic bands is an important characteristic of blazars. It is a powerful tool to better understand the geometry of the emitting regions and the different underlying emission mechanisms \citep[e.g.,][]{2003A&A...400..487C}. Blazars belong to the jetted subclass of AGNs, and have one of their relativistic jets pointing close to the observer. The nonthermal relativistic jet emission from blazar is, therefore, highly Doppler boosted. The Doppler boosted nonthermal jet emission generally swamps out the thermal emission from the accretion disk (AD) feeding the supermassive black hole. Hence, for blazars, jet based theoretical models are mostly used to explain variations at all observable frequencies occurring on different timescales \citep[e.g.][]{1985ApJ...298..114M,1995ARA&A..33..163W}. However, when the blazar is in its very low state and the jet may be very weak, the thermal emission from the AD can dominate over the nonthermal jet emission. In such cases, the microvariability in the light curves of blazars might be attributed to instabilities in the AD or eclipsing of the hotspots on the AD \citep[e.g.,][]{1993ApJ...406..420M,1993ApJ...411..602C}.

There are several possible explanations for brightness variations on STV/LTV timescales in the light curves of blazars. Propagation of a shock in the relativistic jet that accelerates electrons to high energies followed by subsequent cooling via synchrotron and inverse-Compton radiations, the basic shock-in-jet model, \cite[e.g.][]{1985ApJ...298..114M}, can nicely explain most large flares.  Many other changes can be understood to arise from small variations in the viewing angle, and hence in the Doppler factor, caused by either wiggling or helical jets or the motion of the most intense emitting region on a roughly helical trajectory within the jet \citep[e.g.,][]{1992A&A...255...59C, 1992A&A...259..109G, 1999A&A...347...30V}. Variations on shorter (IDV) timescales can be explained by the turbulence expected in a relativistic plasma jet \citep[e.g.,][]{2014ApJ...780...87M, 2015JApA...36..255C, 2016ApJ...820...12P}. 

Observations at optical bands have suggested that the amplitude of flux variations on IDV timescales in the LCs of HBLs is significantly lower than that in the LCs of LBLs \citep[e.g.,][]{1998A&A...329..853H,1999A&AS..135..477R,2011MNRAS.416..101G}. Our optical IDV results are consistent with this conclusion, as we found almost no statistically significant optical IDV in these many light curves of the TeV HBL 1ES 0806$+$524. \cite{1999A&AS..135..477R} suggested that the difference in the optical microvariability behavior of HBLs and LBLs could be due to the presence of stronger magnetic fields in HBLs. An axial magnetic field $B$ can prevent the formation of Kelvin-Helmholtz instabilities in the bases of jets if its value exceeds the critical value $B_c$ given by \citep{1995Ap&SS.234...49R}
\begin{equation}
B_c = \big[4\pi n_e m_e c^2(\Gamma^2 - 1)\big]^{1/2} \Gamma^{-1},
\end{equation}
where $n_e$ is the local electron density, $m_e$ is the electron rest mass, and $\Gamma$ is the bulk Lorentz factor of the jet flow.
In HBLs, such as in 1ES 0806$+$524, higher magnetic fields ($B > B_c$) would prevent the formation of small-scale structures that could produce microvariability or IDV in their optical light curves.

On LTV timescales, we found variations at all three optical frequencies; the amplitude of variability increases with increasing frequency. Our result is consistent with earlier studies \citep[e.g.][]{2003A&A...397..565P,2015MNRAS.452.4263G}. This can be explained by the shock-in-jet model: the electrons accelerated at the shock front suffer energy losses via synchrotron radiation \citep{1985ApJ...298..114M}. The highest energy electrons radiate faster than the lower energy electrons and so emit radiation within a thin region behind the shock front. The thickness of the emitting region increases as the frequency decreases. Therefore, at higher frequencies the variability amplitude is higher and the variability timescale is shorter.

In blazars, flux variations are often linked to changes in color. Blazars generally show one of two different color behaviors: a bluer-when-brighter (BWB) or a redder-when-brighter (RWB) trend. BL Lac objects usually follow a BWB trend, while an RWB trend is generally observed in FSRQs \citep[e.g.,][]{2012MNRAS.425.3002G,2015MNRAS.452.4263G}. However, no clear trend in color variation has also been reported in some blazars \citep[e.g.,][]{2009ApJ...694..174B,2009ApJS..185..511P}. It even has been found that the same source can simultaneously evince both these color behaviors when different pairs of filters are used to compute the colors \cite{2011MNRAS.418.1640W}. In the present work, the HBL 1ES 0806$+$524 shows a BWB trend on LTV timescales. The BWB chromatism may indicate the presence of two components contributing to the overall optical emission: one variable component with a flatter slope and the other stable component with a steeper slope \citep{2004A&A...419...25F.
}

\section{Summary} \label{sec:summ}
In this study, we have presented the optical ($VRI$) photometric observations of the TeV HBL 1ES 0806$+$524 taken with seven ground-based telescopes on 153 nights between 2011 January 25 and 2019 April 12. The results we obtained are summarized below: \\
\\
$\bullet$ No clearly significant intraday variation was observed in any optical band light curve during 17 of 18 nights and it was very weak on that remaining night. \\
\\
$\bullet$ We found no variations in $V-R$ color with time on IDV timescales. \\ 
\\
$\bullet$ On LTV timescales, flux variations were seen in all three optical ($V$, $R$, and $I$) bands. The variability amplitudes were 84.3 \%, 81.6 \%, and 76.5 \% in $V$, $R$ and $I$ bands, respectively. \\ 
\\
$\bullet$ There were modest color variations over this 8 year span of observations. Only the $R-I$ color showed a detectable trend over time, while there were significant correlations of $V-I$, and $R-I$ colors with brightness. \\
\\
$\bullet$ No systematic temporal variation was found in the spectral index, while the optical spectral index was significantly correlated with the $V-$band  magnitude.  \\
\\
This study concludes that, at optical frequencies, the TeV HBL 1ES 0806$+$524 shows almost no statistically significant variation on IDV timescales, while on longer timescales, it exhibits large flux variations and follows the general BWB trend seen for HBLs. It has rather steep optical spectra, with a mean spectral index of 0.639$\pm$0.002, consistent with synchrotron emission dominating the optical light and with the presence of relatively strong axial magnetic fields in the jet.
\\
\\ We thank the anonymous referee for the useful comments and suggestions that help us in improving our manuscript. SOK acknowledges financial support by Shota Rustaveli National Science Foundation of Georgia under contract PHDF-18-354. The Abastumani team acknowledges financial support by the Shota Rustaveli National Science Foundation of Georgia under contract FR/217554/16. JZ acknowledges support from National Natural Science Foundation of China (Grant No. 11973050). The Bulgarian team acknowledges support from the Bulgarian NSF through grants DN08-1/2016, DN 18-13/2017, DN 18-10/2017, KP-06-H28/3 (2018) and KP-06-PN38/1 (2019). GD and OV acknowledge observing grant support from the Institute of Astronomy and Rozhen National Astronomical Observatory, Bulgarian Academy of Sciences, through the bilateral joint research project ``Study of ICRF radio-sources and fast variable astronomical objects'' (for the period 2017-2019, head - G. Damljanovic). This work is a part of the project No. 176011 ``Dynamics and kinematics of celestial bodies and systems'', No. 176004 ``Stellar physics'' and No. 17021 ``Visible and invisible matter in nearby galaxies: theory and observations'' supported by the Ministry of Education, Science and Technological Development of the Republic of Serbia. 

\software{IRAF (\url{http://iraf.net}), DAOPHOT \RNum{2} \citep{1987PASP...99..191S,1992ASPC...25..297S}, python 2.7 (\url{http://www.python.org})}.

\bibliographystyle{aasjournal}
\bibliography{master}

\begin{thebibliography}{}
\expandafter\ifx\csname natexlab\endcsname\relax\def\natexlab#1{#1}\fi
\providecommand{\url}[1]{\href{#1}{#1}}
\providecommand{\dodoi}[1]{doi:~\href{http://doi.org/#1}{\nolinkurl{#1}}}
\providecommand{\doeprint}[1]{\href{http://ascl.net/#1}{\nolinkurl{http://ascl.net/#1}}}
\providecommand{\doarXiv}[1]{\href{https://arxiv.org/abs/#1}{\nolinkurl{https://arxiv.org/abs/#1}}}

\bibitem[{{Abdo} {et~al.}(2010){Abdo}, {Ackermann}, {Agudo}, {Ajello}, {Aller},
  {Aller}, {Angelakis}, {Arkharov}, {Axelsson}, {Bach}, \&
  et~al.}]{2010ApJ...716...30A}
{Abdo}, A.~A., {Ackermann}, M., {Agudo}, I., {et~al.} 2010, \apj, 716, 30,
  \dodoi{10.1088/0004-637X/716/1/30}

\bibitem[{{Acciari} {et~al.}(2009){Acciari}, {Aliu}, {Arlen}, {Bautista},
  {Beilicke}, {Benbow}, {B{\"o}ttcher}, {Bradbury}, {Buckley}, {Bugaev},
  {Butt}, {Byrum}, {Cannon}, {Celik}, {Cesarini}, {Chow}, {Ciupik}, {Cogan},
  {Colin}, {Cui}, {Dickherber}, {Duke}, {Ergin}, {Falcone}, {Fegan}, {Finley},
  {Finnegan}, {Fortin}, {Fortson}, {Furniss}, {Gall}, {Gibbs}, {Gillanders},
  {Grube}, {Guenette}, {Gyuk}, {Hanna}, {Hays}, {Holder}, {Horan}, {Hui},
  {Humensky}, {Imran}, {Kaaret}, {Karlsson}, {Kertzman}, {Kieda}, {Kildea},
  {Konopelko}, {Krawczynski}, {Krennrich}, {Lang}, {LeBohec}, {Maier},
  {McCann}, {McCutcheon}, {Millis}, {Moriarty}, {Mukherjee}, {Nagai}, {Ong},
  {Otte}, {Pandel}, {Perkins}, {Petry}, {Pohl}, {Quinn}, {Ragan}, {Reyes},
  {Reynolds}, {Roache}, {Rose}, {Schroedter}, {Sembroski}, {Smith}, {Steele},
  {Swordy}, {Theiling}, {Toner}, {Valcarcel}, {Varlotta}, {Vassiliev},
  {Wagner}, {Wakely}, {Ward}, {Weekes}, {Weinstein}, {White}, {Williams},
  {Wissel}, {Wood}, \& {Zitzer}}]{2009ApJ...690L.126A}
{Acciari}, V., {Aliu}, E., {Arlen}, T., {et~al.} 2009, \apjl, 690, L126,
  \dodoi{10.1088/0004-637X/690/2/L126}

\bibitem[{{Aleksi{\'c}} {et~al.}(2015){Aleksi{\'c}}, {Ansoldi}, {Antonelli},
  {Antoranz}, {Babic}, {Bangale}, {Barrio}, {Becerra Gonz{\'a}lez}, {Bednarek},
  {Bernardini}, {Biasuzzi}, {Biland}, {Blanch}, {Bonnefoy}, {Bonnoli},
  {Borracci}, {Bretz}, {Carmona}, {Carosi}, {Colin}, {Colombo}, {Contreras},
  {Cortina}, {Covino}, {Da Vela}, {Dazzi}, {De Angelis}, {De Caneva}, {De
  Lotto}, {de O{\~n}a Wilhelmi}, {Delgado Mendez}, {Di Pierro}, {Dominis
  Prester}, {Dorner}, {Doro}, {Einecke}, {Eisenacher}, {Elsaesser},
  {Fern{\'a}ndez-Barral}, {Fidalgo}, {Fonseca}, {Font}, {Frantzen}, {Fruck},
  {Galindo}, {Garc{\'\i}a L{\'o}pez}, {Garczarczyk}, {Garrido Terrats}, {Gaug},
  {Godinovi{\'c}}, {Gonz{\'a}lez Mu{\~n}oz}, {Gozzini}, {Hadasch}, {Hanabata},
  {Hayashida}, {Herrera}, {Hose}, {Hrupec}, {Idec}, {Kadenius}, {Kellermann},
  {Knoetig}, {Kodani}, {Konno}, {Krause}, {Kubo}, {Kushida}, {La Barbera},
  {Lelas}, {Lewandowska}, {Lindfors}, {Lombardi}, {Longo}, {L{\'o}pez},
  {L{\'o}pez-Coto}, {L{\'o}pez-Oramas}, {Lorenz}, {Lozano}, {Makariev},
  {Mallot}, {Maneva}, {Mannheim}, {Maraschi}, {Marcote}, {Mariotti},
  {Mart{\'\i}nez}, {Mazin}, {Menzel}, {Mirand a}, {Mirzoyan}, {Moralejo},
  {Munar-Adrover}, {Nakajima}, {Neustroev}, {Niedzwiecki}, {Nievas Rosillo},
  {Nilsson}, {Nishijima}, {Noda}, {Orito}, {Overkemping}, {Paiano},
  {Palatiello}, {Paneque}, {Paoletti}, {Paredes}, {Paredes-Fortuny}, {Persic},
  {Poutanen}, {Prada Moroni}, {Prandini}, {Puljak}, {Reinthal}, {Rhode},
  {Rib{\'o}}, {Rico}, {Rodriguez Garcia}, {Saito}, {Saito}, {Satalecka},
  {Scalzotto}, {Scapin}, {Schultz}, {Schweizer}, {Shore}, {Sillanp{\"a}{\"a}},
  {Sitarek}, {Snidaric}, {Sobczynska}, {Stamerra}, {Steinbring}, {Strzys},
  {Takalo}, {Takami}, {Tavecchio}, {Temnikov}, {Terzi{\'c}}, {Tescaro},
  {Teshima}, {Thaele}, {Torres}, {Toyama}, {Treves}, {Vogler}, {Will}, {Zanin},
  {Berger}, {Buson}, {D'Ammand o}, {Gasparrini}, {Hovatta}, {Max-Moerbeck},
  {Readhead}, \& {Richards}}]{2015MNRAS.451..739A}
{Aleksi{\'c}}, J., {Ansoldi}, S., {Antonelli}, L.~A., {et~al.} 2015, \mnras,
  451, 739, \dodoi{10.1093/mnras/stv895}

\bibitem[{{Bade} {et~al.}(1998){Bade}, {Beckmann}, {Douglas}, {Barthel},
  {Engels}, {Cordis}, {Nass}, \& {Voges}}]{1998A&A...334..459B}
{Bade}, N., {Beckmann}, V., {Douglas}, N.~G., {et~al.} 1998, \aap, 334, 459

\bibitem[{{B{\"o}ttcher}(2007)}]{2007Ap&SS.307...69B}
{B{\"o}ttcher}, M. 2007, \apss, 307, 69, \dodoi{10.1007/s10509-006-9213-x}

\bibitem[{{B{\"o}ttcher} {et~al.}(2009){B{\"o}ttcher}, {Fultz}, {Aller},
  {Aller}, {Apodaca}, {Arkharov}, {Bach}, {Bachev}, {Berdyugin}, {Buemi},
  {Calcidese}, {Carosati}, {Charlot}, {Ciprini}, {di Paola}, {Dolci},
  {Efimova}, {Scurrats}, {Frasca}, {Gupta}, {Hagen-Thorn}, {Heidt}, {Hiriart},
  {Konstantinova}, {Kopatskaya}, {L{\"a}hteenm{\"a}ki}, {Lanteri}, {Larionov},
  {LeCampion}, {Leto}, {Lindfors}, {Mihov}, {Nieppola}, {Nilsson}, {Ovcharov},
  {P{\"a}{\"a}kk{\"o}nen}, {Pasanen}, {Ragozzine}, {Raiteri}, {Ros}, {Sadun},
  {Sanchez}, {Semkov}, {Sorcia}, {Strigachev}, {Takalo}, {Tornikoski},
  {Trigilio}, {Umana}, {Valcheva}, {Villata}, {Volvach}, {Wu}, \&
  {Zhou}}]{2009ApJ...694..174B}
{B{\"o}ttcher}, M., {Fultz}, K., {Aller}, H.~D., {et~al.} 2009, \apj, 694, 174,
  \dodoi{10.1088/0004-637X/694/1/174}

\bibitem[{{Calafut} \& {Wiita}(2015)}]{2015JApA...36..255C}
{Calafut}, V., \& {Wiita}, P.~J. 2015, Journal of Astrophysics and Astronomy,
  36, 255, \dodoi{10.1007/s12036-015-9324-2}

\bibitem[{{Camenzind} \& {Krockenberger}(1992)}]{1992A&A...255...59C}
{Camenzind}, M., \& {Krockenberger}, M. 1992, \aap, 255, 59

\bibitem[{{Carini}(1990)}]{1990PhDT........11C}
{Carini}, M.~T. 1990, PhD thesis, Georgia State University, Atlanta.

\bibitem[{{Chakrabarti} \& {Wiita}(1993)}]{1993ApJ...411..602C}
{Chakrabarti}, S.~K., \& {Wiita}, P.~J. 1993, \apj, 411, 602,
  \dodoi{10.1086/172862}

\bibitem[{{Ciprini} {et~al.}(2003){Ciprini}, {Tosti}, {Raiteri}, {Villata},
  {Ibrahimov}, {Nucciarelli}, \& {Lanteri}}]{2003A&A...400..487C}
{Ciprini}, S., {Tosti}, G., {Raiteri}, C.~M., {et~al.} 2003, \aap, 400, 487,
  \dodoi{10.1051/0004-6361:20030045}

\bibitem[{{de Diego}(2014)}]{2014AJ....148...93D}
{de Diego}, J.~A. 2014, \aj, 148, 93, \dodoi{10.1088/0004-6256/148/5/93}

\bibitem[{{de Diego} {et~al.}(2015){de Diego}, {Polednikova}, {Bongiovanni},
  {P{\'e}rez Garc{\'{\i}}a}, {De Leo}, {Verdugo}, \&
  {Cepa}}]{2015AJ....150...44D}
{de Diego}, J.~A., {Polednikova}, J., {Bongiovanni}, A., {et~al.} 2015, \aj,
  150, 44, \dodoi{10.1088/0004-6256/150/2/44}

\bibitem[{{Fossati} {et~al.}(1998){Fossati}, {Maraschi}, {Celotti}, {Comastri},
  \& {Ghisellini}}]{1998MNRAS.299..433F}
{Fossati}, G., {Maraschi}, L., {Celotti}, A., {Comastri}, A., \& {Ghisellini},
  G. 1998, \mnras, 299, 433, \dodoi{10.1046/j.1365-8711.1998.01828.x}

\bibitem[{{Gaur} {et~al.}(2012{\natexlab{a}}){Gaur}, {Gupta}, {Strigachev},
  {Bachev}, {Semkov}, {Wiita}, {Peneva}, {Boeva}, {Kacharov}, {Mihov}, \&
  {Ovcharov}}]{2012MNRAS.420.3147G}
{Gaur}, H., {Gupta}, A.~C., {Strigachev}, A., {et~al.} 2012{\natexlab{a}},
  \mnras, 420, 3147, \dodoi{10.1111/j.1365-2966.2011.20243.x}

\bibitem[{{Gaur} {et~al.}(2012{\natexlab{b}}){Gaur}, {Gupta}, {Strigachev},
  {Bachev}, {Semkov}, {Wiita}, {Peneva}, {Boeva}, {Slavcheva-Mihova}, {Mihov},
  {Latev}, \& {Pandey}}]{2012MNRAS.425.3002G}
---. 2012{\natexlab{b}}, \mnras, 425, 3002,
  \dodoi{10.1111/j.1365-2966.2012.21583.x}

\bibitem[{{Gaur} {et~al.}(2015){Gaur}, {Gupta}, {Bachev}, {Strigachev},
  {Semkov}, {B{\"o}ttcher}, {Wiita}, {de Diego}, {Gu}, {Guo}, {Joshi}, {Mihov},
  {Palma}, {Peneva}, {Rajasingam}, \& {Slavcheva-Mihova}}]{2015MNRAS.452.4263G}
{Gaur}, H., {Gupta}, A.~C., {Bachev}, R., {et~al.} 2015, \mnras, 452, 4263,
  \dodoi{10.1093/mnras/stv1556}

\bibitem[{{Gopal-Krishna} {et~al.}(2011){Gopal-Krishna}, {Goyal}, {Joshi},
  {Karthick}, {Sagar}, {Wiita}, {Anupama}, \& {Sahu}}]{2011MNRAS.416..101G}
{Gopal-Krishna}, {Goyal}, A., {Joshi}, S., {et~al.} 2011, \mnras, 416, 101,
  \dodoi{10.1111/j.1365-2966.2011.19014.x}

\bibitem[{{Gopal-Krishna} \& {Wiita}(1992)}]{1992A&A...259..109G}
{Gopal-Krishna}, \& {Wiita}, P.~J. 1992, \aap, 259, 109

\bibitem[{{Gupta} {et~al.}(2004){Gupta}, {Banerjee}, {Ashok}, \&
  {Joshi}}]{2004A&A...422..505G}
{Gupta}, A.~C., {Banerjee}, D.~P.~K., {Ashok}, N.~M., \& {Joshi}, U.~C. 2004,
  \aap, 422, 505, \dodoi{10.1051/0004-6361:20040306}

\bibitem[{{Gupta} {et~al.}(2016){Gupta}, {Agarwal}, {Bhagwan}, {Strigachev},
  {Bachev}, {Semkov}, {Gaur}, {Damljanovic}, {Vince}, \&
  {Wiita}}]{2016MNRAS.458.1127G}
{Gupta}, A.~C., {Agarwal}, A., {Bhagwan}, J., {et~al.} 2016, \mnras, 458, 1127,
  \dodoi{10.1093/mnras/stw377}

\bibitem[{{Gupta} {et~al.}(2019){Gupta}, {Gaur}, {Wiita}, {Pand ey},
  {Kushwaha}, {Hu}, {Kurtanidze}, {Semkov}, {Damljanovic}, {Goyal}, {Uemura},
  {Darriba}, {Chen}, {Vince}, {Gu}, {Zhang}, {Bachev}, {Chanishvili}, {Itoh},
  {Kawabata}, {Kurtanidze}, {Nakaoka}, {Nikolashvili}, {Stawarz}, \&
  {Strigachev}}]{2019AJ....157...95G}
{Gupta}, A.~C., {Gaur}, H., {Wiita}, P.~J., {et~al.} 2019, \aj, 157, 95,
  \dodoi{10.3847/1538-3881/aafe7d}

\bibitem[{{Heidt} \& {Wagner}(1996)}]{1996A&A...305...42H}
{Heidt}, J., \& {Wagner}, S.~J. 1996, \aap, 305, 42

\bibitem[{{Heidt} \& {Wagner}(1998)}]{1998A&A...329..853H}
---. 1998, \aap, 329, 853

\bibitem[{{Kshama} {et~al.}(2017){Kshama}, {Paliya}, \&
  {Stalin}}]{2017MNRAS.466.2679K}
{Kshama}, S.~K., {Paliya}, V.~S., \& {Stalin}, C.~S. 2017, \mnras, 466, 2679,
  \dodoi{10.1093/mnras/stw3317}

\bibitem[{{Kurtanidze} {et~al.}(2009){Kurtanidze}, {Tetradze}, {Richter},
  {Nikolashvili}, {Kimeridze}, \& {Sigua}}]{2009ASPC..408..266K}
{Kurtanidze}, O.~M., {Tetradze}, S.~D., {Richter}, G.~M., {et~al.} 2009, in
  Astronomical Society of the Pacific Conference Series, Vol. 408, The
  Starburst-AGN Connection, ed. W.~{Wang}, Z.~{Yang}, Z.~{Luo}, \& Z.~{Chen},
  266

\bibitem[{{Man} {et~al.}(2014){Man}, {Zhang}, {Wu}, {Zhou}, \&
  {Yuan}}]{2014AJ....148..110M}
{Man}, Z., {Zhang}, X., {Wu}, J., {Zhou}, X., \& {Yuan}, Q. 2014, \aj, 148,
  110, \dodoi{10.1088/0004-6256/148/6/110}

\bibitem[{{Mangalam} \& {Wiita}(1993)}]{1993ApJ...406..420M}
{Mangalam}, A.~V., \& {Wiita}, P.~J. 1993, \apj, 406, 420,
  \dodoi{10.1086/172453}

\bibitem[{{Marcha} {et~al.}(1996){Marcha}, {Browne}, {Impey}, \&
  {Smith}}]{1996MNRAS.281..425M}
{Marcha}, M.~J.~M., {Browne}, I.~W.~A., {Impey}, C.~D., \& {Smith}, P.~S. 1996,
  \mnras, 281, 425, \dodoi{10.1093/mnras/281.2.425}

\bibitem[{{Marscher}(2014)}]{2014ApJ...780...87M}
{Marscher}, A.~P. 2014, \apj, 780, 87, \dodoi{10.1088/0004-637X/780/1/87}

\bibitem[{{Marscher} \& {Gear}(1985)}]{1985ApJ...298..114M}
{Marscher}, A.~P., \& {Gear}, W.~K. 1985, \apj, 298, 114,
  \dodoi{10.1086/163592}

\bibitem[{{Miller} {et~al.}(1989){Miller}, {Carini}, \&
  {Goodrich}}]{1989Natur.337..627M}
{Miller}, H.~R., {Carini}, M.~T., \& {Goodrich}, B.~D. 1989, \nat, 337, 627,
  \dodoi{10.1038/337627a0}

\bibitem[{{Padovani} {et~al.}(2017){Padovani}, {Alexander}, {Assef}, {De
  Marco}, {Giommi}, {Hickox}, {Richards}, {Smol{\v{c}}i{\'c}},
  {Hatziminaoglou}, {Mainieri}, \& {Salvato}}]{2017A&ARv..25....2P}
{Padovani}, P., {Alexander}, D.~M., {Assef}, R.~J., {et~al.} 2017, \aapr, 25,
  2, \dodoi{10.1007/s00159-017-0102-9}

\bibitem[{{Pandey} {et~al.}(2019){Pandey}, {Gupta}, {Wiita}, \&
  {Tiwari}}]{2019ApJ...871..192P}
{Pandey}, A., {Gupta}, A.~C., {Wiita}, P.~J., \& {Tiwari}, S.~N. 2019, \apj,
  871, 192, \dodoi{10.3847/1538-4357/aaf974}

\bibitem[{{Papadakis} {et~al.}(2003){Papadakis}, {Boumis}, {Samaritakis}, \&
  {Papamastorakis}}]{2003A&A...397..565P}
{Papadakis}, I.~E., {Boumis}, P., {Samaritakis}, V., \& {Papamastorakis}, J.
  2003, \aap, 397, 565, \dodoi{10.1051/0004-6361:20021581}

\bibitem[{{Polednikova} {et~al.}(2016){Polednikova}, {Ederoclite}, {de Diego},
  {Cepa}, {Gonz{\'a}lez-Serrano}, {Bongiovanni}, {Oteo}, {Garc{\'{\i}}a},
  {P{\'e}rez-Mart{\'{\i}}nez}, {Pintos-Castro}, {Ram{\'o}n-P{\'e}rez}, \&
  {S{\'a}nchez-Portal}}]{2016MNRAS.460.3950P}
{Polednikova}, J., {Ederoclite}, A., {de Diego}, J.~A., {et~al.} 2016, \mnras,
  460, 3950, \dodoi{10.1093/mnras/stw1252}

\bibitem[{{Pollack} {et~al.}(2016){Pollack}, {Pauls}, \&
  {Wiita}}]{2016ApJ...820...12P}
{Pollack}, M., {Pauls}, D., \& {Wiita}, P.~J. 2016, \apj, 820, 12,
  \dodoi{10.3847/0004-637X/820/1/12}

\bibitem[{{Poon} {et~al.}(2009){Poon}, {Fan}, \& {Fu}}]{2009ApJS..185..511P}
{Poon}, H., {Fan}, J.~H., \& {Fu}, J.~N. 2009, \apjs, 185, 511,
  \dodoi{10.1088/0067-0049/185/2/511}

\bibitem[{{Rees}(1984)}]{1984ARA&A..22..471R}
{Rees}, M.~J. 1984, \araa, 22, 471, \dodoi{10.1146/annurev.aa.22.090184.002351}

\bibitem[{{Reinthal} {et~al.}(2012){Reinthal}, {Lindfors}, {Mazin}, {Nilsson},
  {Takalo}, {Sillanp{\"a}{\"a}}, {Berdyugin}, \& {MAGIC
  Collaboration}}]{2012JPhCS.355a2013R}
{Reinthal}, R., {Lindfors}, E.~J., {Mazin}, D., {et~al.} 2012, in Journal of
  Physics Conference Series, Vol. 355, Journal of Physics Conference Series,
  012013, \dodoi{10.1088/1742-6596/355/1/012013}

\bibitem[{{Romero}(1995)}]{1995Ap&SS.234...49R}
{Romero}, G.~E. 1995, \apss, 234, 49, \dodoi{10.1007/BF00627281}

\bibitem[{{Romero} {et~al.}(1999){Romero}, {Cellone}, \&
  {Combi}}]{1999A&AS..135..477R}
{Romero}, G.~E., {Cellone}, S.~A., \& {Combi}, J.~A. 1999, \aaps, 135, 477,
  \dodoi{10.1051/aas:1999184}

\bibitem[{{Schachter} {et~al.}(1993){Schachter}, {Stocke}, {Perlman}, {Elvis},
  {Remillard}, {Granados}, {Luu}, {Huchra}, {Humphreys}, {Urry}, \&
  {Wallin}}]{1993ApJ...412..541S}
{Schachter}, J.~F., {Stocke}, J.~T., {Perlman}, E., {et~al.} 1993, \apj, 412,
  541, \dodoi{10.1086/172942}

\bibitem[{{Stetson}(1987)}]{1987PASP...99..191S}
{Stetson}, P.~B. 1987, \pasp, 99, 191, \dodoi{10.1086/131977}

\bibitem[{{Stetson}(1992)}]{1992ASPC...25..297S}
{Stetson}, P.~B. 1992, in Astronomical Society of the Pacific Conference
  Series, Vol.~25, Astronomical Data Analysis Software and Systems I, ed. D.~M.
  {Worrall}, C.~{Biemesderfer}, \& J.~{Barnes}, 297

\bibitem[{{Stocke} {et~al.}(1991){Stocke}, {Morris}, {Gioia}, {Maccacaro},
  {Schild}, {Wolter}, {Fleming}, \& {Henry}}]{1991ApJS...76..813S}
{Stocke}, J.~T., {Morris}, S.~L., {Gioia}, I.~M., {et~al.} 1991, \apjs, 76,
  813, \dodoi{10.1086/191582}

\bibitem[{{Urry} \& {Padovani}(1995)}]{1995PASP..107..803U}
{Urry}, C.~M., \& {Padovani}, P. 1995, \pasp, 107, 803, \dodoi{10.1086/133630}

\bibitem[{{Villata} \& {Raiteri}(1999)}]{1999A&A...347...30V}
{Villata}, M., \& {Raiteri}, C.~M. 1999, \aap, 347, 30

\bibitem[{{Wagner} \& {Witzel}(1995)}]{1995ARA&A..33..163W}
{Wagner}, S.~J., \& {Witzel}, A. 1995, \araa, 33, 163,
  \dodoi{10.1146/annurev.aa.33.090195.001115}

\bibitem[{{Wierzcholska} {et~al.}(2015){Wierzcholska}, {Ostrowski}, {Stawarz},
  {Wagner}, \& {Hauser}}]{2015A&A...573A..69W}
{Wierzcholska}, A., {Ostrowski}, M., {Stawarz}, {\L}., {Wagner}, S., \&
  {Hauser}, M. 2015, \aap, 573, A69, \dodoi{10.1051/0004-6361/201423967}

\bibitem[{{Wu} {et~al.}(2012){Wu}, {B{\"o}ttcher}, {Zhou}, {He}, {Ma}, \&
  {Jiang}}]{2012AJ....143..108W}
{Wu}, J., {B{\"o}ttcher}, M., {Zhou}, X., {et~al.} 2012, \aj, 143, 108,
  \dodoi{10.1088/0004-6256/143/5/108}

\bibitem[{{Wu} {et~al.}(2011){Wu}, {Zhou}, {Ma}, \&
  {Jiang}}]{2011MNRAS.418.1640W}
{Wu}, J., {Zhou}, X., {Ma}, J., \& {Jiang}, Z. 2011, \mnras, 418, 1640,
  \dodoi{10.1111/j.1365-2966.2011.19565.x}

\end{thebibliography}
\end{document}